\documentclass[aps,prl,nobalancelastpage,twocolumn,superscriptaddress,reprint]{revtex4-2}
\usepackage{ulem}
\usepackage{hyperref}
\usepackage{graphicx}
\usepackage{tabularx}
\usepackage{booktabs}
\usepackage{caption}
\usepackage{subcaption}
\usepackage{bm, physics}
\usepackage[skip=4.5pt]{caption}

\hypersetup{
    colorlinks=true,
    linkcolor=blue,
    filecolor=gray,      
    urlcolor=blue,
    citecolor=blue,
}
\usepackage{comment}

\usepackage{ragged2e}
\newcommand{\BLUE}[1]{{\color{black} #1}}

\begin{document}

\title{Restoring the Surface Magnetic Gap in MnBi$_2$Te$_4$}

\author{Ce \surname{Bian}}
\affiliation{State Key Laboratory of Low Dimensional Quantum Physics and Department of Physics, Tsinghua University, Beijing, 100084, China}

\author{Hengxin \surname{Tan}}
\email{hxtan@sjtu.edu.cn}
\affiliation{Key Laboratory of Artificial Structures and Quantum Control (Ministry of Education), School of Physics and Astronomy, Shanghai Jiao Tong University, Shanghai, China}

\author{Wenhui \surname{Duan}}
\affiliation{State Key Laboratory of Low Dimensional Quantum Physics and Department of Physics, Tsinghua University, Beijing, 100084, China}
\affiliation{Institute for Advanced Study, Tsinghua University, Beijing 100084, China}
\affiliation{Frontier Science Center for Quantum Information, Beijing, China}


\begin{abstract}
A widespread experimental realization of quantized anomalous transport in the intrinsic magnetic topological insulator MnBi$_2$Te$_4$ is hindered by its elusive surface magnetic gap. Uncovering the origin of the gapless states is thus essential for accessing its topological properties. Here we show that surface defects lower the electrostatic potential, drive topological surface states into subsurface layers, suppress exchange interactions, thereby closing the gap.
Tuning the surface electrostatic potential via external electric fields or interfacial fields in van der Waals heterostructures restores the expected gap and enables control of its topology. This is confirmed by model calculations and validated in defective MnBi$_2$Te$_4$ films interfaced with polar insulators, explaining the enhanced quantum anomalous Hall effect under AlO$_x$ capping observed in recent experiments.
\BLUE{Our theory identifies electrostatically driven surface-state delocalization as a competitive origin of gap suppression and proposes displacement-field engineering for robust quantized transport.}
\end{abstract}

\maketitle

\textit{Introduction.}
Since the discovery of the quantum Hall effect in the 1980s \cite{Klitzing1980New}, achieving quantized charge transport without external magnetic fields has remained a central goal in condensed matter physics \cite{haldane1988model,hasan2010colloquium,qi2011topological}. Despite extensive efforts, the experimental realization of the quantum anomalous Hall effect (QAHE) has proven difficult, with only limited success \cite{chang2023RMP}. The first observation, in Cr-doped (Bi,Sb)$_2$Te$_3$, occurred at an ultra low temperature of 30 mK \cite{chang2013experimental}, far below the threshold for practical applications. Moreover, magnetic doping complicates material synthesis and further hinders the robust realization of QAHE. These challenges underscore the urgent need to raise the quantization temperature and ease the experimental realization of QAHE.

Among various theoretical proposals, the magnetic topological insulator MnBi$_2$Te$_4$ has emerged as a promising platform for realizing high-temperature QAHE and axion insulators \cite{li2019intrinsic,PhysRevLett.122.206401}. First-principles calculations predict that this layered compound hosts an A-type antiferromagnetic order and a bulk topological band gap of 0.2 eV, along with a sizable surface magnetic gap of 90 meV on its natural cleavage terminations \cite{li2019intrinsic,PhysRevLett.122.206401}. Its van der Waals (vdW) structure further facilitates device fabrication and tunability. The successful synthesis of MnBi$_2$Te$_4$ \cite{gong2019experimental,otrokov2019prediction} generated considerable excitement \cite{he2020mnbi2te4,zhao2021routes,wang2023on,li2023progress,vyazovskaya2025intrinsic}. However, despite early reports claiming evidence of the surface gap \cite{otrokov2019prediction,lee2019spin,shikin2021sample,vidal2019surface,Ji2021Detection}, most ARPES measurements revealed gapless surface states  \cite{vidal2019topological,li2019dirac,chen2019topological,Swatek2020gapless,XU20202086,hu2020universal,
Nevola2020Coexistence}, even though surface ferromagnetism appeared robust \cite{Nevola2020Coexistence,Sass2020robust}. This discrepancy poses a major obstacle to realizing quantized transport in MnBi$_2$Te$_4$ and likely accounts for the rarity of experimental observations of QAHE \cite{deng2020quantum,ge2020high,lian2025antiferromagnetic,wang2025towards,guo2025quantized} and axion insulators \cite{liu2020robust,qiu2025observation} in this intriguing material.

To resolve the discrepancy between theory and experiment, several mechanisms have been proposed, including surface–bulk hybridization \cite{ma2020hybridization}, frustrated surface magnetism \cite{chen2019topological,Swatek2020gapless,li2019magnetically}, structure reconstruction or degradation \cite{hou2020te,yuan2020electronic,Sattar2025surface}, surface vdW gap expansion \cite{shikin2020nature,wang2023three}, and defect-induced cancellation of magnetic exchange interactions \cite{tan2023distinct}.
Nevertheless, a definitive understanding of the underlying physics has yet to be achieved.
Given the prevalence of intrinsic defects in MnBi$_2$Te$_4$ samples \cite{liu2021site,lai2021defect,huang2020native,Wimmer2021AM,murakami2019realization}, defects are likely a major factor in suppressing the surface magnetic gap \cite{shikin2021sample,AFMp2006516,wu2023irremovable,garnica2022native,lupke2025defect,guo2025quantized}. We emphasize that the defect-induced gap reduction mechanism can reconcile current experimental inconsistencies \cite{tan2023distinct}.
\BLUE{However, the microscopic mechanism underlying defect influence on surface states remains to be fully understood.}
Experimentally, identifying practical routes to restore the surface gap and enable high-temperature quantized transport remains a pressing challenge.

\begin{figure*}
\includegraphics[width= 0.94\textwidth]{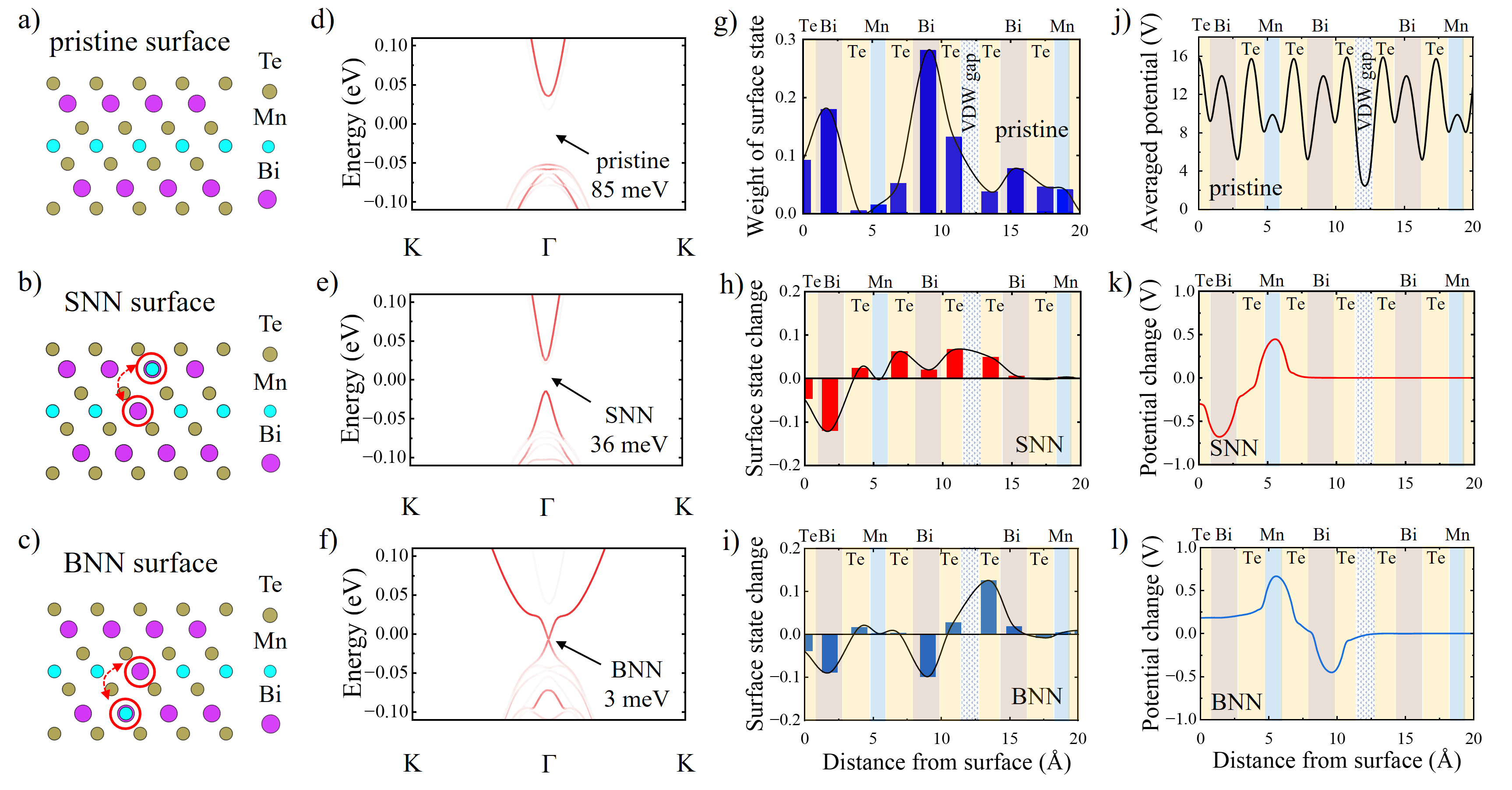}
\caption{ \justifying a–c) Top-SL schematics of the pristine, SNN-defective, and BNN-defective surfaces. SNN/BNN refers to the nearest-neighbor Mn-Bi co-antisites near the surface/bulk side of the top SL. Defects are marked with red circles in b-c).
d-f) Surface-projected band structures. Darker (lighter) red corresponds to higher (lower) surface SL weight.
g) Spatial distribution of the valence band maximum (VBM) of the pristine surface.
h,i) SNN and BNN defect-induced changes of the VBM distribution relative to the pristine surface in g).
j) Electrostatic potential of the pristine surface.
k,l) SNN and BNN defect-induced changes in electrostatic potential relative to the pristine surface in j).
}
\label{fig1}
\end{figure*}

In this work, we reveal that surface defects reduce the surface electrostatic potential of MnBi$_2$Te$_4$, leading to substantial penetration of topological surface states into subsurface layers.
To counteract this effect, we propose tuning the surface potential via electric fields, thereby reopening the expected surface topological gap. Two strategies are explored. The first involves applying an external electric field, which, as confirmed by \BLUE{model and first-principles} calculations, significantly enhances the gap \BLUE{and controls its topology} on defective surfaces. The second exploits interfacial electric fields by coupling defective MnBi$_2$Te$_4$ to trivial polar insulators. This approach is validated through calculations for a realistic heterostructure with the ferroelectric In$_2$Te$_3$ \cite{ding2017prediction}, where a sizable surface gap is recovered. Inspired by recent reports of QAHE enhancement in AlO$_x$-capped MnBi$_2$Te$_4$ \cite{lian2025antiferromagnetic}, we further examine AlO$_x$ and common capping layers and \BLUE{demonstrate the general applicability of our theory}.

\textit{Defect-induced surface gap reduction.}
Previous studies have revealed the prevalence of intrinsic defects in MnBi$_2$Te$_4$ samples \cite{
murakami2019realization,liu2021site,huang2020native,hou2020te,lai2021defect,Wimmer2021AM}.
Here, we focus on the most common and energetically favorable type$-$Mn–Bi co-antisite defects \cite{AFMp2006516,tan2023distinct,wu2023irremovable,murakami2019realization}. We consider two configurations SNN and BNN for such defects: nearest-neighbor Mn–Bi antisite pairs within the surface septuple layer (SL), where ``S" (``B") indicates that the Bi atom lies closer to the surface (bulk) side [see Fig. \ref{fig1}(a-c)].
To evaluate their impact on surface states, we employed slab models incorporating these defects. Technical details and more results are provided in the Supplemental Material (SM) \cite{SM}.

As shown in Fig. \ref{fig1}(d), the pristine MnBi$_2$Te$_4$ surface exhibits a magnetic gap of 85 meV. The corresponding atomic-layer-resolved distribution of the topological surface states is presented in Fig. \ref{fig1}(g). Upon introducing Mn–Bi co-antisites, the surface gap is notably reduced to 36 meV for SNN [Fig. \ref{fig1}(e)] and merely 3 meV for BNN [Fig. \ref{fig1}(f)] surface. Similar behavior is observed for other defects (see SFIG. 1 in SM \cite{SM}).
Concomitantly, the surface states become delocalized from the topmost SL and penetrate deeper into the subsurface, as evidenced by the reduced wavefunction weight on the surface atomic layers and the enhanced contribution from atoms close to the vdW gap [Fig. \ref{fig1}(h,i), see also SFIG. 4 in SM \cite{SM}]. Such redistribution leads to destructive interference in magnetic exchange interactions across the top two SLs \cite{tan2023distinct}.
Notably, defect and sample preparation-induced relocation of surface states has recently been demonstrated in experiment \cite{li2024fabrication,lupke2025defect}.

This inward relocation closely correlates with defect-induced changes in the surface electrostatic potential [Fig. \ref{fig1}(k,l)], relative to the pristine surface [Fig. \ref{fig1}(j)], \BLUE{which scale with defect concentration (see SFIGs. 3\&5 in SM \cite{SM})}.
Specifically, substituting Bi with Mn lowers the potential in the Bi layer, while replacing Mn with Bi raises the Mn-layer potential. These trends reflect the intrinsic potential hierarchy in the pristine surface [Fig. \ref{fig1}(j)], where Bi (Mn) layers possess higher (lower) electrostatic potentials.
Notably, the reduced potential in the Bi layer has a more pronounced effect on the surface states, as surface states are primarily derived from Bi/Te orbitals [See Fig. \ref{fig1}(g-i)].
Thus, we conclude that the suppressed electrostatic potential of the Bi layer is the driving force behind the subsurface relocation of the topological states and the resulting gap reduction.

Having identified the microscopic origin of the surface gap suppression, a highly watched question arises: how can the gap be restored to enable high-temperature quantized transport? Defect minimization is a straightforward but challenging route \cite{guo2025quantized}. We propose an alternative strategy that circumvents sample quality limitations: raise the surface electrostatic potential to drag the topological states back to the surface, thereby mitigating magnetic exchange cancellation. This can be achieved either through an external vertical electric field or via interfacial electric fields by coupling MnBi$_2$Te$_4$ to polar insulators.

\begin{figure}[h]
\includegraphics[width= 0.5\textwidth]{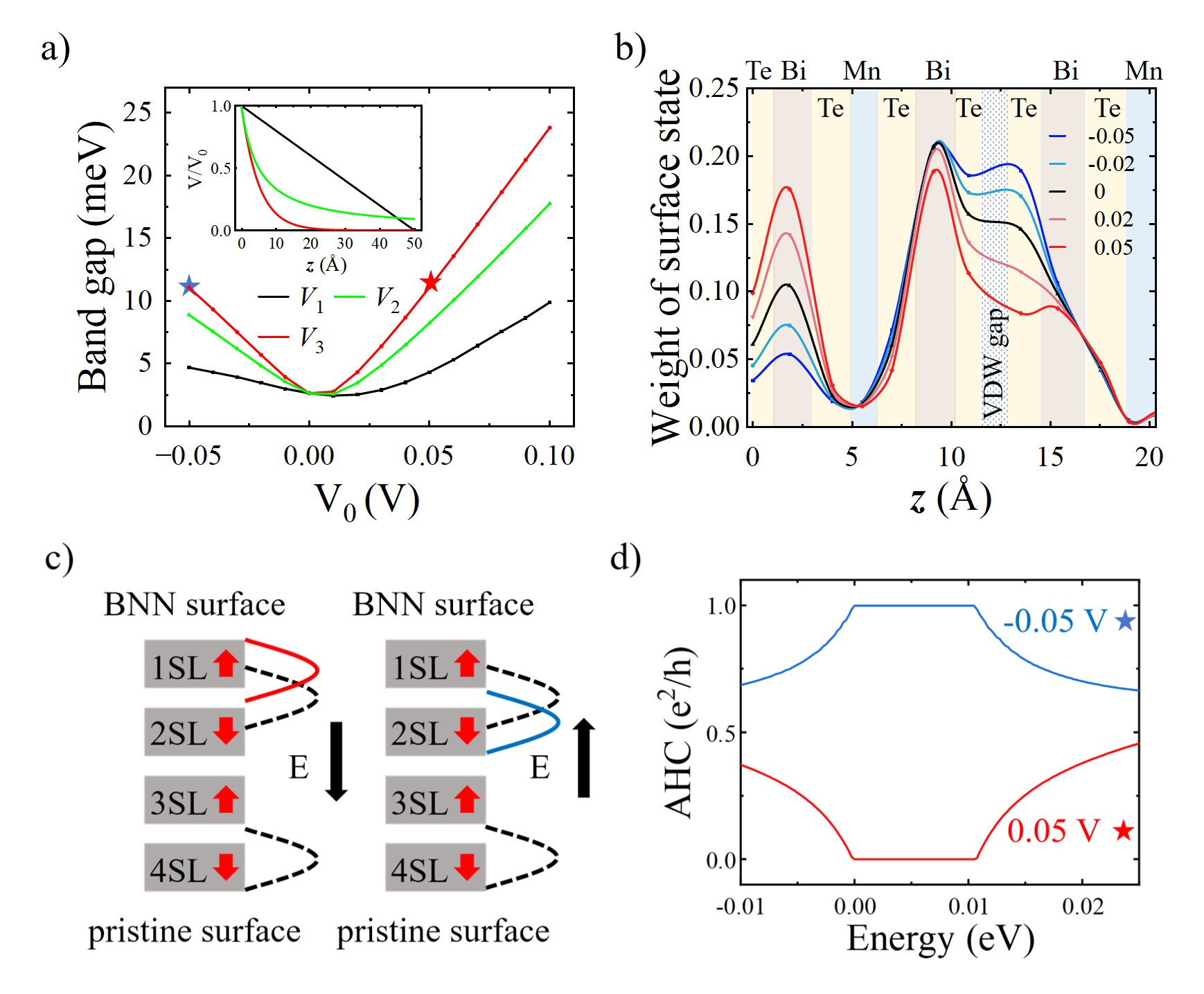}
\caption{ \justifying
a) BNN surface magnetic gap under three representative electric fields. Inset: spatial profiles of electric fields with $\lambda_1 = 50$ \AA, $\lambda_2 = 5$ \AA, and $\lambda_3 = 5$ \AA.
b) Distribution of the \BLUE{valence band top} under the exponentially decaying field $V_3$ with varying $V_0$.
c) Schematic of electric-field effects on surface states: dashed black curve, before applying electric fields; solid red (blue) curve, after applying a downward (upward) field $E$ (bold black arrow). Red arrows indicate magnetization in each SL.
d) AHC of the four-SL slab under $V_3$ with $V_0 = \pm0.05$ V. The valence band top is set to zero energy. Red and blue stars mark corresponding positions in a). 
}
\label{fig2}
\end{figure}

\textit{Electric field-induced gap restoration \BLUE{and topology switching.}}
We first examine the feasibility of reopening the magnetic gap on a defective MnBi$_2$Te$_4$ surface using external electric fields. We consider three representative forms of electric potential (or electric field) along the surface normal direction ($z$-axis) that may occur in experiments \cite{S1}:
\begin{itemize}
\item Uniform electric field: $V_1(z) = V_0 - V_0 \cdot z/\lambda_1$
\item Point-charge-like field: $V_2(z) = V_0 \cdot \frac{\lambda_2}{z + \lambda_2}$
\item Exponentially decaying field: $V_3(z) = V_0 \cdot e^{-z/\lambda_3}$
\end{itemize}
Here, $\lambda_i$ ($i=1,2,3$) is a screening parameter controlling the decay length of the potential$-$the smaller $\lambda_i$, the faster the decay. We set the topmost Te layer at $z = 0$, with increasing $z$ denoting depth into the material. 
In $V_2$, the virtual $point~charge$ is placed $\lambda_2$ above the surface \cite{S2}.
In all cases, the surface layer is held at the same potential $V_0$, which defines the field direction: $V_0 > 0$ increases the surface potential relative to the bulk, while $V_0 < 0$ decreases it. The spatial profiles of the electric potentials are sketched in the inset of Fig. \ref{fig2}(a).

\BLUE{We exemplify the field effect with a four-SL BNN-defective slab, whose small magnetic gap affords the full tunability of topological surface states by electric fields}. The tight-binding Hamiltonian is constructed via maximally localized Wannier functions \cite{Mostofi2008}, upon which the external electric fields are applied as on-site energy terms (see choice of $\lambda_i$ and band structures SFIG. 6-8 in SM \cite{SM}). Figure \ref{fig2}(a) shows the evolution of the surface magnetic gap as a function of $V_0$.
As it shows, increasing $V_0$ significantly reopens the magnetic gap. For instance, at $V_0 = 0.1$ V, the surface gap increases to 11 meV, 19 meV, and 24 meV under uniform, point-charge-like, and exponentially decaying electric fields, respectively$-$far exceeding the initial 3 meV gap in the BNN-defective surface.
\BLUE{Such surface gap modulations by electric fields, supported by DFT calculations (SFIG. 9 \cite{SM}), are insensitive to defect type and slab thickness [SFIGs. 10-12 \cite{SM}]}.
Faster-decaying fields (e.g., $V_3$) reopen the gap more effectively, likely due to their concentrated impact on the surface and resultant weak bulk-surface coupling.

The atomic layer-resolved distribution of the topological surface state exemplified under $V_3$ in Fig. \ref{fig2}(b) \BLUE{[see also SFIG. 13 \cite{SM}]} confirms that higher surface potentials ($V_0>0$) relocalize surface states towards the outermost SL, which decreases the magnetic effect of the subsurface SL. Conversely, a negative $V_0$ lowers the surface potential, shifting the topological surface states deeper into the second SL. This diminishes the magnetic influence of the top SL and also reopens the gap.
The dependence of the surface-state relocation on electric field direction is sketched in Fig. \ref{fig2}(c).
\BLUE{These observations contrast with the inward surface-state shift and gap suppression predicted for all potentials in relevant regime in the literature \cite{Menshov2022towards}.}

Crucially, the topology of the gap induced by negative $V_0$ differs fundamentally from that induced by positive $V_0$. Figure \ref{fig2}(d) shows the anomalous Hall conductivity (AHC) as a function of energy for the four-SL slab under $V_3$ with $V_0 = \pm0.05$ V. At $V_0 = +0.05$ V, the AHC within the gap is zero, consistent with an even-layered MnBi$_2$Te$_4$ predicted to be an axion insulator, where the half-quantized AHC from opposite surfaces cancel \cite{gu2021spectral} (see layer-resolved AHC in SFIG. 15 \cite{SM}). However, at $V_0 = -0.05$ V, the AHC exhibits a quantized plateau of $\frac{e^2}{\hbar}$, characteristic of an odd-layered MnBi$_2$Te$_4$ as a Chern insulator. \BLUE{The topological nature of the gaps is further demonstrated by their characteristic edge states (SFIG. 14 \cite{SM})}. We note that adopting an odd-layered slab model in calculations would reverse the AHC response \BLUE{(see SFIG. 12 \cite{SM})}.
These results indicate that relocating surface states to the second SL effectively reduces the slab thickness and switches the system between a QAHE and an axion insulator.

\begin{figure}[h]
\centering
\includegraphics[width= 0.48\textwidth]{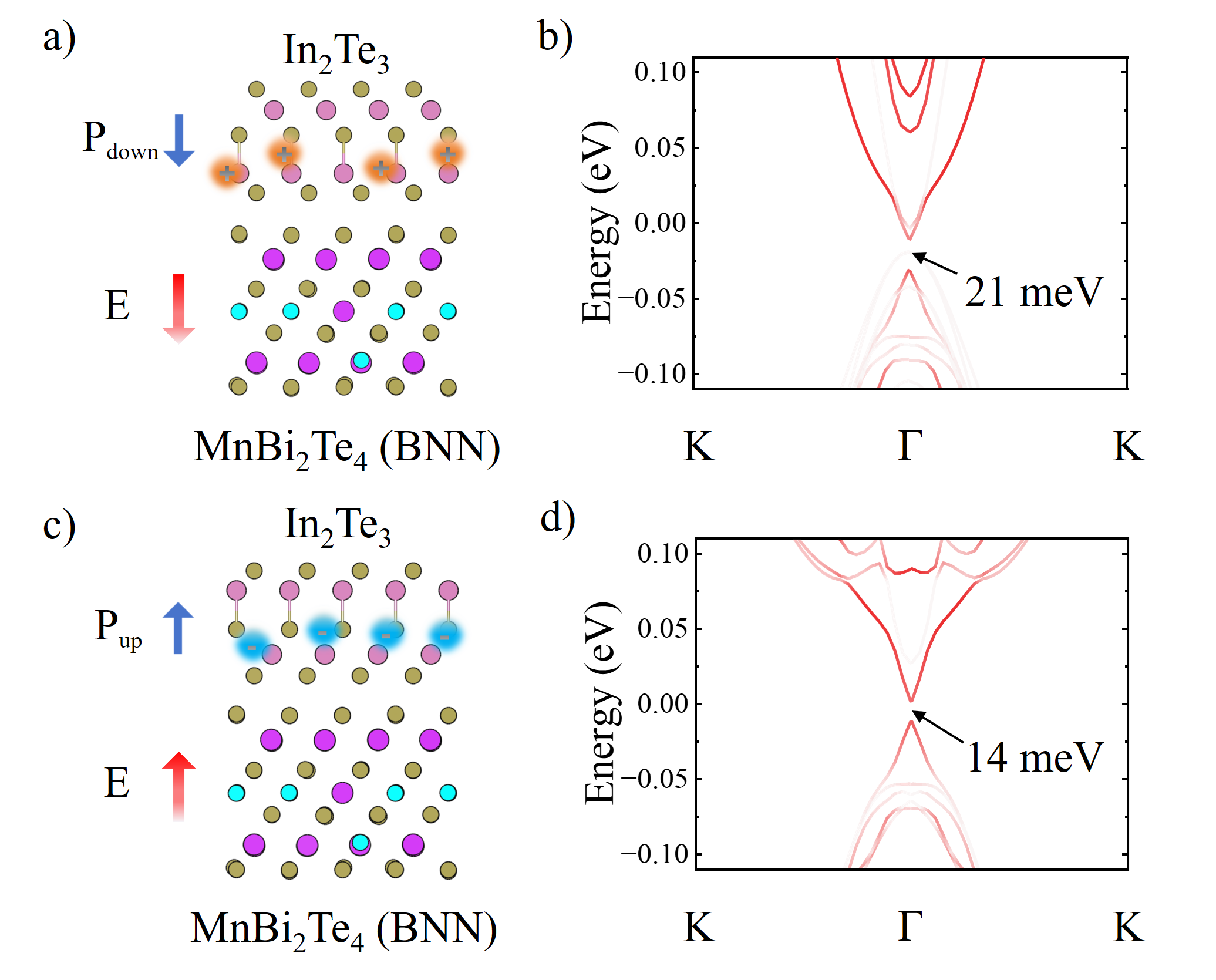}
\caption{\justifying
a) Side view of \BLUE{4-SL} BNN-defective MnBi$_2$Te$_4$ capped with downward-polarized In$_2$Te$_3$ ($P_{down}$, blue arrow), showing interfacial positive charge and induced downward electric field $E$ (red arrow) in MnBi$_2$Te$_4$. Only the topmost SL is displayed.
b) Surface-projected energy dispersion of the structure in a). The intensity of the color red is proportional to the surface SL weight.
c–d) Same as a–b), but with upward-polarized In$_2$Te$_3$ ($P_{up}$), leading to a reverse of the interfacial electric field.
}
\label{fig3}
\end{figure}

The sensitivity of the surface states to the local electrostatic potential modulation offers a viable strategy to tune surface electronic states via electrostatic engineering. In particular, the dependence of the gap topology on the electric field direction underscores the critical importance of field orientation in device design and might find applications in topological field-effect transistors.
Notably, the required field strengths at the surface ($z = 0$) are on the order of $10^7$–$10^8$ V/m for $V_0=0.1$ V, well within the range achievable in modern nanodevices.

\begin{figure}[h]
\centering
\includegraphics[width= 0.5\textwidth]{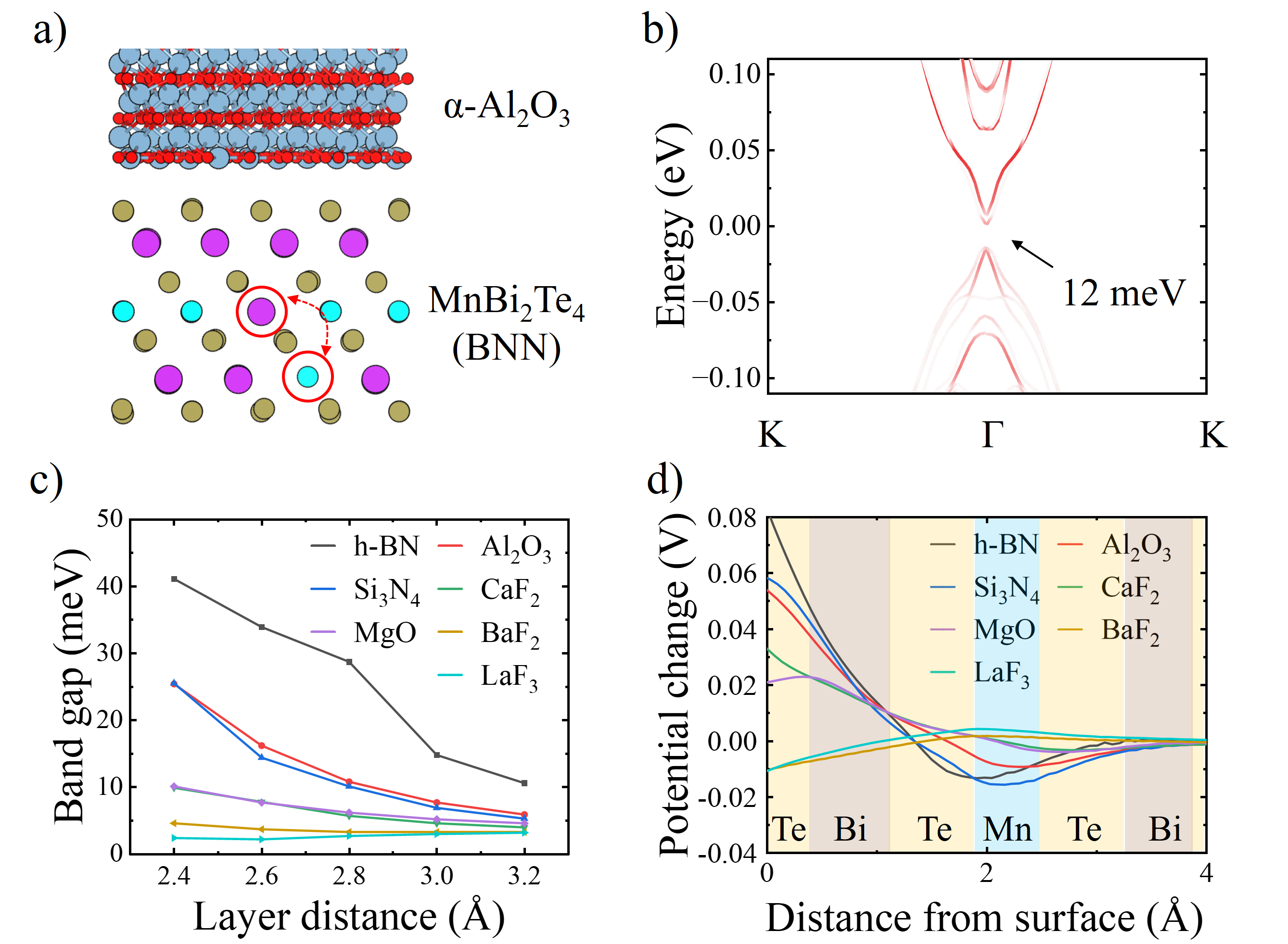}
\caption{ \justifying
a) The \BLUE{4-SL} BNN-defective MnBi$_2$Te$_4$ slab capped with Al$_2$O$_3$, with corresponding band dispersion in b). Darker (lighter) red in b) corresponds to higher (lower) weight of the topmost SL.
c) Band gap evolution with vdW spacing between the surface SL and capping layers, including oxides, nitrides, and fluorides.
d) Electrostatic potential changes (relative to the uncapped defective surface) at the equilibrium vdW spacing.
}
\label{fig4}
\end{figure}

\textit{Interface engineering of the magnetic gap.}
Beyond applying external electric fields, the surface electrostatic potential can also be modulated via interface engineering by placing the material in contact with capping layers. This approach is particularly relevant for experiments, as samples always involve substrates and are often encapsulated between protective layers. We begin by considering a heterostructure composed of the \BLUE{four-SL} BNN-defective surface and a polar insulator possessing a trivial electronic structure. Such a polar layer can tune the electrostatic potential at the interface via its polarization, while its insulating nature avoids the inclusion of any exotic states within the MnBi$_2$Te$_4$ magnetic gap. A suitable candidate is the layered ferroelectric insulator series In$_2$Te$_3$ with a strong out-of-plane polarization \cite{ding2017prediction,li2025ferroelectric}. Indeed, the resulting heterostructures [Fig. \ref{fig3}(a,c)] exhibit much larger magnetic gaps [see Fig. \ref{fig3}(b,d)] than that of the uncapped defective surface$-$demonstrating the efficacy of this strategy.

The mechanism resembles that of an external electric field. Downward polarization of In$_2$Te$_3$ [Fig. \ref{fig3}(a)] accumulates positive charges at the interface, producing a downward electric field in MnBi$_2$Te$_4$ that raises the surface potential and drags the surface states back to the top SL. Conversely, upward polarization [Fig. \ref{fig3}(c)] accumulates negative interface charges, generating an upward field that lowers the surface potential and pushes the surface state into the second SL. Both cases reopen the magnetic gap, but with distinct magnitudes (21 meV vs 14 meV, Fig. \ref{fig3}(b,d)). The larger gap from downward polarization can be attributed to the rapid decay of the interfacial field in MnBi$_2$Te$_4$, which allows it to pull surface states toward the surface more efficiently than an upward field pushes them into the second SL.

Recent experiments have shown that capping MnBi$_2$Te$_4$ with AlO$_x$ significantly promotes the quantum anomalous Hall effect, attributed to the enhanced surface magnetic anisotropy by AlO$_x$ capping \cite{lian2025antiferromagnetic,wang2025towards}. While such anisotropy favors out-of-plane spin alignment, the observed gap reopening likely involves additional factors, as gapless states have also been reported under the surface A-type magnetic order \cite{Nevola2020Coexistence,Sass2020robust}. In particular, these studies point to interfacial electric fields arising from broken inversion symmetry as a possible contributor, similar to the mechanism discussed above.

We demonstrate this scenario with the heterostructure formed by the \BLUE{four-SL} BNN-defective MnBi$_2$Te$_4$ and $\alpha$ phase Al$_2$O$_3$ [Fig. \ref{fig4}(a)]. The band structure [Fig. \ref{fig4}(b)] shows that the surface magnetic gap increases to 12 meV, four times that of the uncapped BNN-defective surface (3 meV). Similar results are found with non-stoichiometric AlO$_x$ (see SFIG. 16 \cite{SM}). We further examined various insulating capping layers used in experiments, including oxides, nitrides, and fluorides \cite{kagerer2020molecular,martini2023hall,Li2025nonvolatile,vishwanath2014molecular,imai2010systematic} (see details in SFIG. 17-19 \cite{SM}). Figure \ref{fig4}(c) shows the dependence of the surface magnetic gap on the vdW spacing between MnBi$_2$Te$_4$ and capping layers. Al$_2$O$_3$ and nitrides (BN, Si$_3$N$_4$) are most effective in reopening the gap. The corresponding surface electrostatic potential changes at equilibrium vdW spacing [Fig. \ref{fig4}(d)] confirm that these cappings substantially raise the potential, whereas others have a minor impact. \BLUE{
The reopened gaps are expected to be topological in nature, although a rigorous verification remains computationally prohibitive at present.}

While the above results highlight the effectiveness of interface engineering in tuning MnBi$_2$Te$_4$ surface states, a key challenge is predicting whether a given capping layer enhances/weakens the surface electrostatic potential. For polar materials such as In$_2$Te$_3$, the interfacial electric field is determined by the interfacial charges and thus predictable. For non-polar materials, it may be inferred from work-function or electron-affinity differences, but such indirect criteria are not fully deterministic. First-principles calculations remain the most reliable approach, while developing simpler yet robust predictors, especially for non-polar cappings, remains a material-specific and open problem.

\textit{Conclusion.}
We have shown that tuning the surface electrostatic potential by external electric fields or interfacial fields in heterostructures \BLUE{effectively reshape the spatial profile of the topological surface-state wavefunction} in MnBi$_2$Te$_4$. Consequently, the long-sought surface magnetic gap is efficiently restored, and its topology can be controlled. These findings not only clarify the origin of the missing surface gap in this highly concerning material, but also suggest practical strategies toward achieving high-temperature quantized transport.


\textbf{Acknowledgments.}
We acknowledge helpful discussions with Chang Liu from Renmin University of China. 
H.T. thanks the sponsorship from Yangyang Development Fund and \BLUE{is supported by NSFC (No. 12574270) and the Science and Technology Commission of Shanghai Municipality (No. 24PJA051).
C.B. ans W.D. are supported by the Quantum Science and Technology-National Science and Technology Major Project (Grant No. 2023ZD0300500), the Fundamental and Interdisciplinary Disciplines Breakthrough Plan of the Ministry of Education of China (Grant No. JYB2025XDXM408), the Beijing Advanced Innovation Center for Future Chip (ICFC), and the National Key Basic Research and Development Program of China (Grant Nos. 2023YFA1406400, 2024YFA1409100).}



\begin{thebibliography}{71}
\bibitem [{S1()}]{S1}%
  \BibitemOpen
  \bibinfo {note} {Due to electrostatic screening, applied electric fields decay in materials, particularly in metallic systems such as topological insulators with metallic surface states. Since the exact decay profile is difficult to determine, we illustrate the principles of restoring the surface magnetic gap by considering two representative decaying fields, $V_2$ and $V_3$, which are potentially relevant to experiments.}\BibitemShut {Stop}%
\bibitem [{S2()}]{S2}%
  \BibitemOpen
  \bibinfo {note} {We place the virtual point charge in $V_2$ according to the screening parameter $\lambda_2$, so that the surface layer potential matches $V_0$ and the electric field $\frac{V_0}{\lambda_2}$ in $V_2$ resembles the field $\frac{V_0}{\lambda_1}$ in $V_1$ and $\frac{V_0}{\lambda_3}$ in $V_3$ at $z=0$.}\BibitemShut {Stop}%
\end{thebibliography}

%

\end{document}


\title{Supplemental Material for ``Restoring the Surface Magnetic Gap in MnBi$_2$Te$_4$''}

\author{Ce \surname{Bian}}
\affiliation{State Key Laboratory of Low Dimensional Quantum Physics and Department of Physics, Tsinghua University, Beijing, 100084, China}

\author{Hengxin \surname{Tan}}
\email{hxtan@sjtu.edu.cn}
\affiliation{Key Laboratory of Artificial Structures and Quantum Control (Ministry of Education), School of Physics and Astronomy, Shanghai Jiao Tong University, Shanghai, China}

\author{Wenhui \surname{Duan}}
\affiliation{State Key Laboratory of Low Dimensional Quantum Physics and Department of Physics, Tsinghua University, Beijing, 100084, China}
\affiliation{Institute for Advanced Study, Tsinghua University, Beijing 100084, China}
\affiliation{Frontier Science Center for Quantum Information, Beijing, China}

\maketitle

\section{Table of content}
\begin{itemize}
  \item \textbf{Methodology}: Calculation method and technique details. 
  
  \item \textbf{SFIG. }\ref{sfig1}: Surface defect configurations and corresponding band structures not contained in the main text.
  
  \BLUE{\item \textbf{SIFG. }\ref{Rtable2}: Total energies of different co-antisite configurations under AFM and Ferri-AFM.}

  \BLUE{\item \textbf{SIFG. }\ref{RTable1}: Magnetic gaps of BNN and SNN surfaces as a function of defect concentrations.}

  \item \textbf{SFIG. }\ref{sfig3}: Spatial distribution of the conduction band minimum of the pristine, SNN-, and BNN-defective surfaces.
  
  \item \textbf{SFIG. }\ref{sfig4}: Electrostatic potential modulation as a function of BNN/SNN defect concentration. 
  
  \item \textbf{SFIG. }\ref{sfig5}: Band structure of the BNN surface under the uniform electric field $V_1$ with varying $V_0$ and $\lambda_1$.
  
  \item \textbf{SFIG. }\ref{sfig6}: Band structure of the BNN surface under the point-charge-like field $V_2$ with varying $V_0$ and $\lambda_2$.
  
  \item \textbf{SFIG. }\ref{sfig7}: Band structure of the BNN surface under the exponentially decaying field $V_3$ with varying $V_0$ and $\lambda_3$.

  \BLUE{\item \textbf{SFIG. }\ref{Rfig2}: Band gaps of the BNN surface under the homogeneous electric field from DFT calculations.}

  \BLUE{\item \textbf{SFIG. }\ref{Rfig4}: Influence of the exponentially decaying field on the 4 SL SNN slab.}
  
  \BLUE{\item \textbf{SFIG. }\ref{Rfig1}: Influence of the exponentially decaying field on a 4 SL slab with a Te vacancy.}

  \BLUE{\item \textbf{SFIG. }\ref{Rfig5}: Influence of the exponentially decaying field on a 5 SL BNN slab.}
  
  \BLUE{\item \textbf{SFIG. }\ref{sfig8}: Spatial distributions of the conduction band minimum of the BNN-defective surface under the exponentially decaying field $V_3$.}

  \BLUE{\item \textbf{SFIG. }\ref{Rfig8}: Edge states of the 4 SL BNN slab at negative and positive $V_0$.} 
  
  \item \textbf{SFIG. }\ref{sfig9}: Layer-resolved AHC of the BNN-defective slab under the exponentially decaying field $V_3$ with $V_0 = \pm 0.05$ V.
  
  \item \textbf{SFIG. }\ref{sfig10}: Electronic structure of the BNN-defective MnBi$_2$Te$_4$ surface capped with the non-stoichiometric alumina AlO$_x$.
  
  \item \textbf{SFIG. }\ref{sfig11}: Slab structures of several commonly used substrate materials mentioned in the main text.

  \BLUE{\item \textbf{SFIG. }\ref{Rtable3}: A summary of strain applied to capping layers in heterostructure calculations.}
  
  \item \textbf{SFIG. }\ref{sfig12}: Band structures of heterostructures formed by the BNN-defective MnBi$_2$Te$_4$ and other substrate materials.

  \item \textbf{SFIG. }\ref{sfig13}: Raw band structures for different configurations, with their surface-projected dispersions presented in the main text.

\end{itemize}

\newpage

\section{methodology}
\subsection{DFT and model calculations}

All first-principles calculations were carried out using the Vienna $ab$-$initio$ Simulation Package (VASP) \cite{Kresse1996Efficient}. Tight-binding models were constructed with Wannier90 \cite{Mostofi2008} and TB-models \cite{gresch2018automated}. The exchange–correlation functional was treated using the revised Perdew–Burke–Ernzerhof (PBE) form of the generalized gradient approximation (GGA) \cite{perdew1996generalized}. A plane-wave cutoff energy of 500 eV was applied throughout. To correct the underestimated band gap of MnBi$_2$Te$_4$, we used the GGA+$U$ approach with $U=5$ eV on Mn-$d$ orbitals. All calculations assumed interlayer antiferromagnetic order with ferromagnetic alignment within each septuple layer (i.e., the A-type AFM configuration, \BLUE{which is the most stable configuration, see SFIG. \ref{Rtable2}}). For co-antisite defects, Mn moments were aligned parallel to the other Mn atoms within the same SL.

For all defect calculations, we used a slab model of \BLUE{four} SLs thick with a $3\times3$ in-plane supercell, \BLUE{unless explicitly stated otherwise}. Defects were introduced only on the top surface, while the bottom surface remained pristine. To reduce computational cost, relaxations of the defective surfaces were restricted to in-plane atomic coordinates, with out-of-plane coordinates ($z$) fixed. All calculations, \BLUE{including structural relaxations}, incorporated spin–orbit coupling. \BLUE{The inclusion of spin–orbit coupling in structure relaxation leads to different magnetic gaps for defective surfaces as compared to Ref. \cite{tan2023distinct}.}

To investigate the effect of external electric fields on defective surfaces (main text Fig. 2), we employed the tight-binding model derived from the Wannier90 software, with the external potential implemented as an on-site energy term. Band structures under various field profiles, with different $V_0$ and $\lambda_i$, are shown in Supplementary SFIGs. \ref{sfig5}–\ref{sfig7}. In the main text, we report results for three representative cases: a uniform field $V_1$ with $\lambda_1$=50 \AA, a point-charge-like field $V_2$ with $\lambda_2$=5 \AA, and an exponentially decaying field $V_3$ with $\lambda_3$=5 \AA. These $\lambda_i$ values were chosen so that: (1) the band gap remains direct at the $\Gamma$ point over the selected $V_0$ range, (2) the surface gap stays within the bulk gap, and (3) the gap magnitude is large. \BLUE{Notice that these $\lambda_i$ values are also justified by VASP calculations with the homogeneous electric fields as shown in SFIG. \ref{Rfig2}.}

To study interfacial electric fields induced by polar insulators (main text Fig. 3), we first considered In$_2$Se$_3$ capping. However, its underestimated band gap in standard first-principles calculations led to artificial metallicity, and its lattice mismatch with MnBi$_2$Te$_4$ is substantial. We therefore used In$_2$Te$_3$, which results in an insulating heterostructure and bears a lattice mismatch below 1\%. 
For a broader exploration of capping layers (main text Fig. 4), we built analogous heterostructures with insulating terminations. Specifically, we used the Al-terminated of the $\alpha$-phase Al$_2$O$_3$, the reconstructed mixed Mg–O surface for MgO, and F-terminations for all fluorides. The exact surface terminations are shown in Supplementary Fig. \ref{sfig11}. Because of computational cost, only the BNN-defective MnBi$_2$Te$_4$–Al$_2$O$_3$ heterostructure was fully relaxed, giving an optimized interlayer vdW spacing of $\sim$2.8 \AA~between the MnBi$_2$Te$_4$ surface and the Al$_2$O$_3$ slab. For other capping layers, each component was relaxed separately, and then combined into heterostructures with the vdW spacing treated as a tunable parameter. In all cases, the in-plane lattice constants were fixed to those of MnBi$_2$Te$_4$, \BLUE{and the resulting lattice strain for capping layers is summarized in SFIG. \ref{Rtable3}}.
All models consisted of a four-SL MnBi$_2$Te$_4$ slab (in-plane supercell $3\times3$) with a single pair of BNN co-antisite defects.

For non-stoichiometric AlO$_x$ ($x$ variable) \cite{JEURGENS2002313}, motivated by recent experiments reporting $x>1.5$, we inserted random numbers of O atoms into the Al$_2$O$_3$ slab and fully relaxed the resulting sub-slabs. All BNN-defective MnBi$_2$Te$_4$–AlO$_x$ heterostructures exhibited larger band gaps than the uncapped BNN-defective surface, without a regular dependence on $x$. This may suggest that the surface structure, rather than the exact stoichiometry of AlO$_x$, plays the primary role in tuning the magnetic gap.

\subsection{AHC calculations}

The anomalous Hall conductivity (AHC) $\sigma_{xy}$ was calculated using the Kubo formula on a dense $201\times 201$ $k$-mesh. To ensure convergence, the mesh was further refined by a factor of five in the vicinity of the $\Gamma$ point, where the Berry curvature exhibits strong variations. To obtain the layer-resolved AHC $\sigma_{xy}^{(\ell)}$ for the slab, we followed the approach developed in Refs. \onlinecite{Varnava2018Surfaces,Rauch2018Geometric}, where $\sigma_{xy}^{(\ell)}$ is expressed in terms of the layer-resolved Chern number $C_\ell$ (also referred to as the local Chern marker in the literature):
\[
\sigma_{xy}^{(\ell)} \;=\; \frac{e^2}{h}\,C_{\ell}, \quad
C_{\ell} \;=\; -\frac{4\pi}{A}\,\mathrm{Im}\,\frac{1}{N_k}
\sum_{\mathbf k}\sum_{v v' c}
X_{v c}\,Y^{\dagger}_{v' c}\,\rho^{(\ell)}_{v' v}.
\]
Here
\[
\rho^{(\ell)}_{v' v} \;=\; \bra{\psi_{v'k}} L_\ell \ket{\psi_{vk}},\quad
X_{vc} \;=\; \frac{(v_x)_{vc}}{E_c-E_v},\quad
Y_{vc} \;=\; \frac{(v_y)_{vc}}{E_c-E_v},\quad
(v_{x/y})_{vc} \;=\; \bra{\psi_{vk}}  \partial_{k_{x/y}} H  \ket{\psi_{ck}}.
\]
$\rho^{(\ell)}$ is the matrix representation of the wave-function projector $L_\ell$ onto layer $\ell$. $\psi_{vk}$ and $\psi_{ck}$ denote the eigenvectors of valence and conduction bands at $k$, respectively, and $v/v'$, and $c$ are the corresponding band indices. $X_{vc}$ and $Y_{vc}$ are velocity ($v_x$ and $v_y$) matrix elements, evaluated between valence and conduction states. $x/y$ denotes the direction of the velocity. $E_{v/c}$ are the band eigenvalues, and $A$ is the in-plane unit-cell area.
By writing $X$, $Y$, and $\rho^{(\ell)}$ in matrix form (notice that $X/Y$ are not necessarily square matrices while $\rho^{(\ell)}$ is a square matrix), $C_\ell$ can be simplified as,
\[
\sum_{v v' c} X_{v c}\,Y^{\dagger}_{v' c}\,\rho^{(\ell)}_{v' v}
= \sum_{v v'} \Big(\sum_{c} X_{v c}\,Y^{\dagger}_{v' c}\,\Big) \rho^{(\ell)}_{v' v}
= \sum_{v v'} \big( X\,Y^{\dagger} \big)_{v v'} \rho^{(\ell)}_{v' v}
= \mathrm{Tr} \!\big[XY^{\dagger}\,\rho^{(\ell)}\big].
\]
Thus,
\[
C_{\ell} \;=\; -\frac{4\pi}{A}\,\mathrm{Im}\,\frac{1}{N_k}
\sum_{\mathbf k}\mathrm{Tr}\!\big[X\,Y^{\dagger}\,\rho^{(\ell)}\big].
\]
For a uniform $k$-mesh, the factor 1/$N_k$, where $N_k$ is the total number of $k$ points, provides the correct normalization. In the case of our locally refined $k$-point sampling, it is replaced by the weight $w_{\mathbf{k}}$, defined as the fractional area of $k$-space associated with each sampling point, with $\sum_{\mathbf k} w_{\mathbf{k}} = 1$. The final expression is then
\[
C_{\ell}
= -\frac{4\pi}{A}\,\sum_{\mathbf k} w_{\mathbf k}\;
\Big[\,\mathrm{Im}\,\mathrm{Tr}\!\big(XY^{\dagger}\rho^{(\ell)}\big)\Big].
\]
Notice that summing over all layers $\sum_\ell \sigma_{xy}^{(\ell)}$ exactly reproduces the total intrinsic AHC. Results for the four-SL BNN slab under the exponentially decaying field $V_3$ are shown in SFIG. \ref{sfig9}.

\newpage


\begin{sfigure*}
\centering
\includegraphics[width= 1\textwidth]{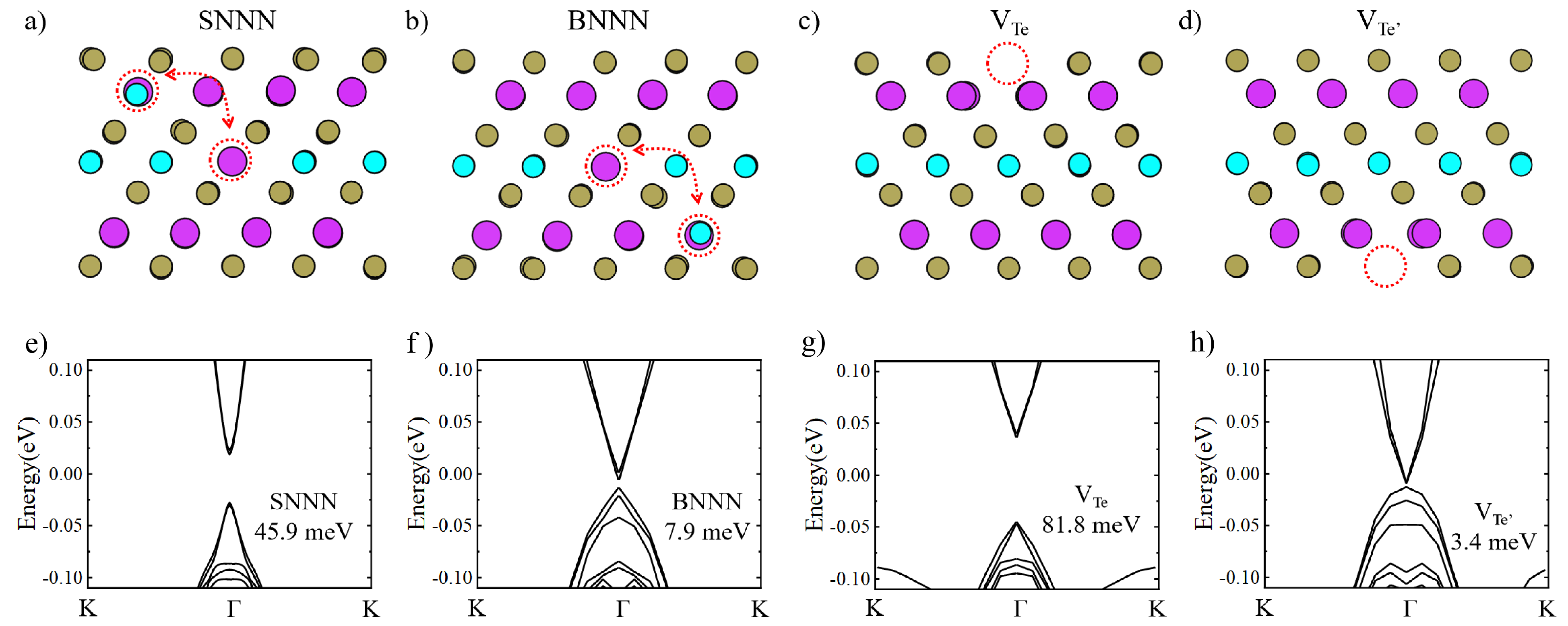}
\caption{\justifying a-d) Schematic illustration of the defective surface in a four-SL slab model where only the topmost SL is shown. The upper edge of the shown SL corresponds to the surface. Defects, highlighted with red circles, include next-nearest-neighbor Mn–Bi antisite pairs (SNNN, BNNN, similar to SNN and BNN in the main text), and Te vacancy (V$_{Te}$, V$_{Te^{'}}$).
e-h) Corresponding band structures of defect configurations in a-d). The in-plane cell size is 3$\times$3. 
}
\label{sfig1}
\end{sfigure*}

\begin{sfigure*}
\centering
\includegraphics[width=1\textwidth]{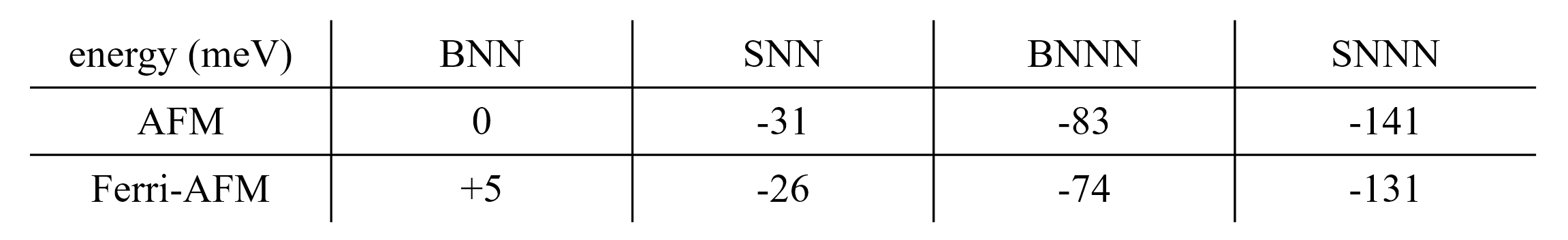}
\caption{\justifying \BLUE{Total energies of different defect configurations in the A-type AFM and Ferri-AFM phases (with Mn$_\mathrm{Bi}$ moments antiparallel to other Mn atoms within the same SL). Energies are referenced to the AFM BNN surface (in meV/cell). SNNN and BNNN are more stable than SNN and BNN, and the Ferri-AFM phase is always less stable than AFM, consistent with Ref. \cite{tan2023distinct}. Notice that SOC inclusion in surface relaxation affects the absolute energy differences.
}
}
\label{Rtable2}
\end{sfigure*}

\begin{sfigure*}
\centering
\includegraphics[width=1\textwidth]{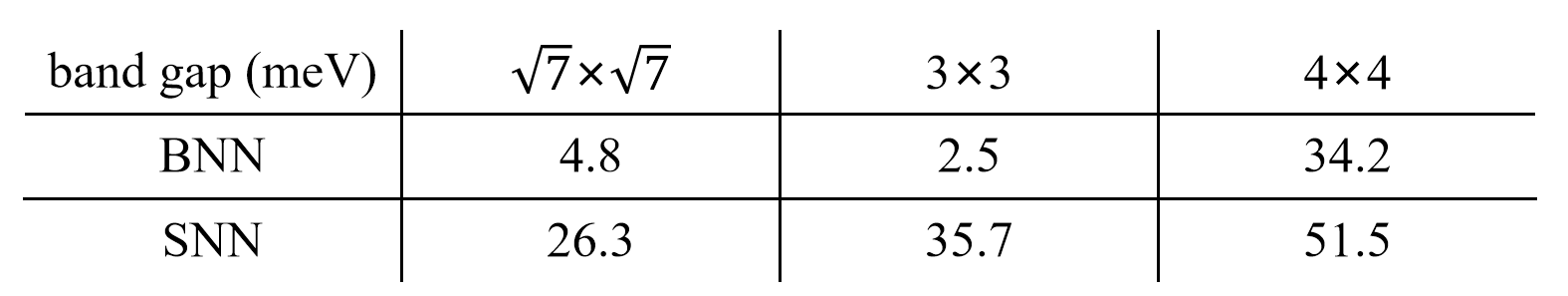}
\caption{\justifying \BLUE{Magnetic gaps of the SNN and BNN defective surfaces as a function of defect concentration. Only one co-antisite pair is present in each in-plane supercell. Smaller supercells correspond to higher defect concentrations, generally leading to smaller magnetic gaps.}
}
\label{RTable1}
\end{sfigure*}

\begin{sfigure*}
\centering
\includegraphics[width= 1.00\textwidth]{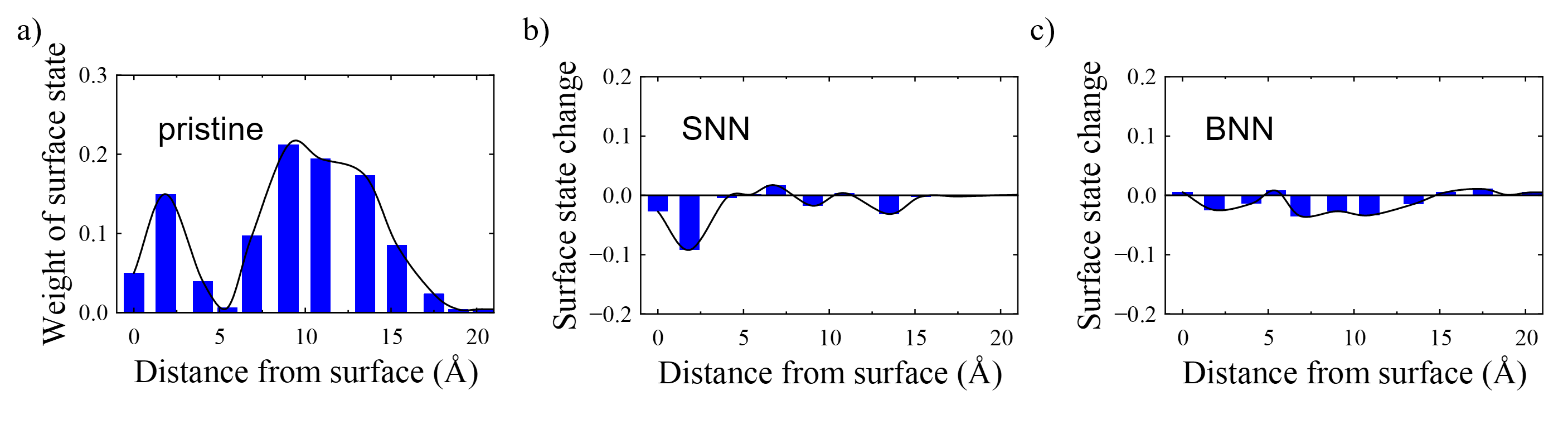}
\caption{\justifying a) Spatial distribution of the conduction band minimum (CBM) of the pristine surface. b,c) SNN and BNN defect-induced changes of the CBM distribution relative to the pristine surface in a).
These results are complementary to Fig. 1(g-i).}
\label{sfig3}
\end{sfigure*}

\begin{sfigure*}
\centering
\includegraphics[width= 1.00\textwidth]{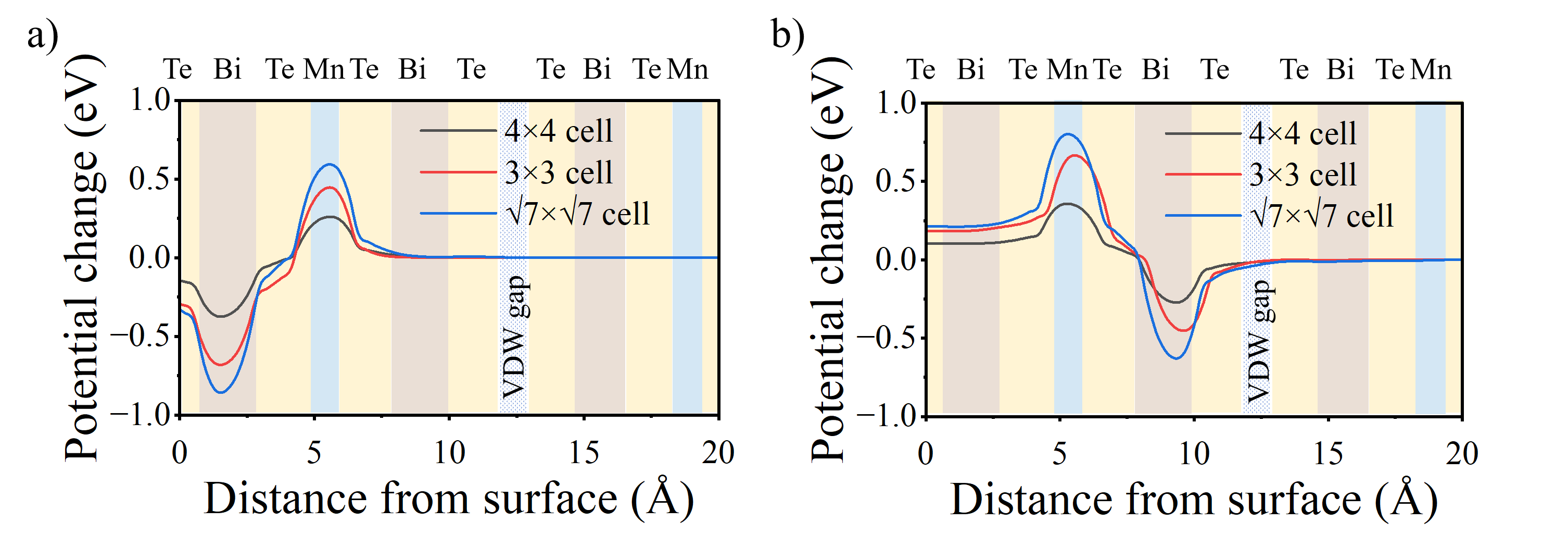}
\caption{\justifying Electrostatic potential variations near the MnBi$_2$Te$_4$ surface induced by a single pair of antisite defects in different supercellss: (a) SNN-defective surface; (b) BNN-defective surface. Higher defect concentration (i.e., smaller cell size) leads to stronger potential modulation.
}
\label{sfig4}
\end{sfigure*}

\begin{sfigure*}
    \centering
    \includegraphics[width= 1.00\textwidth]{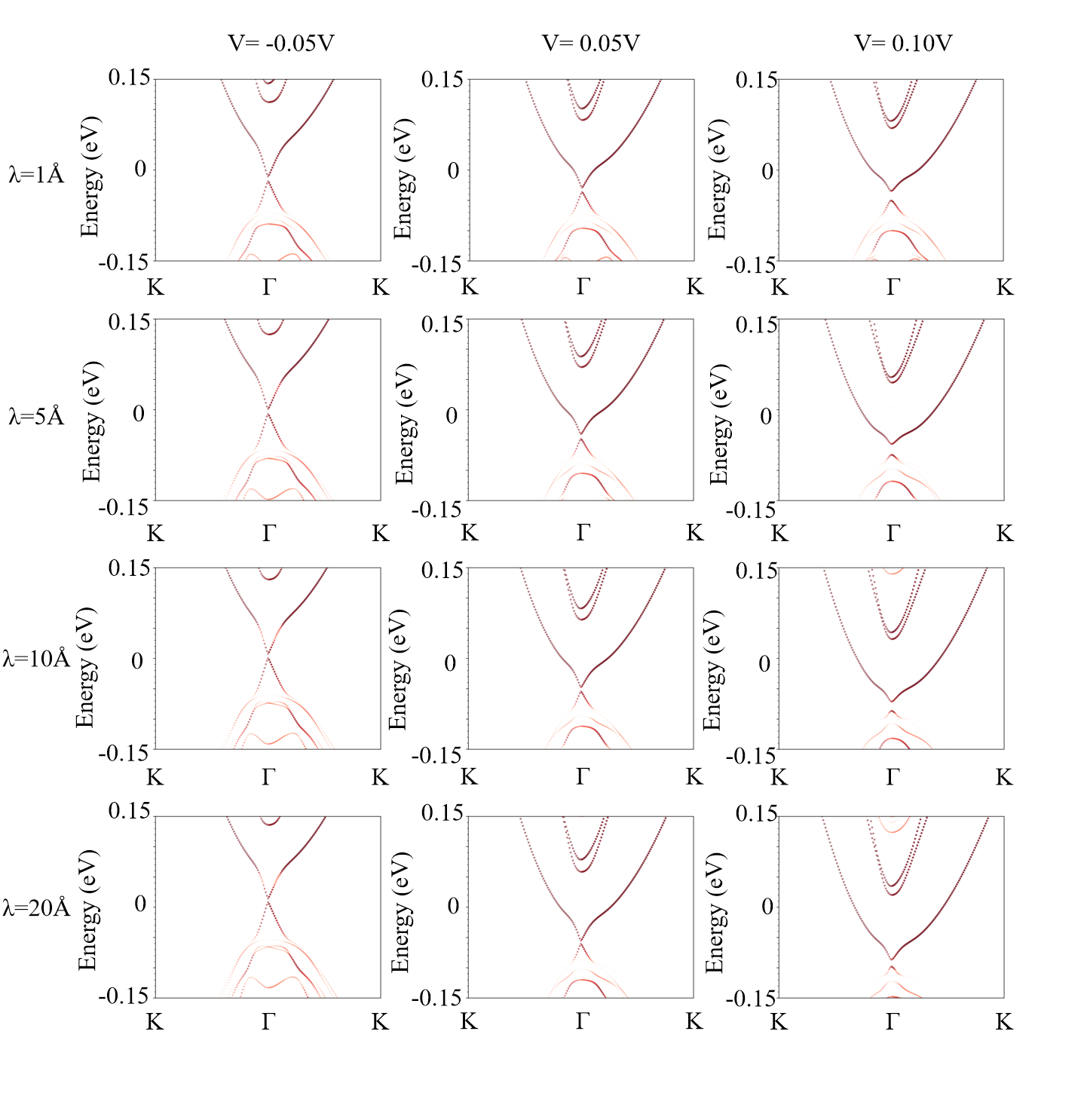}
    \caption{\justifying Band structures of the BNN defect surface calculated with the uniform electric field $V_1$. The intensity of the color red is proportional to the contribution from the surface. Different rows and columns correspond to different values of $V_0$ and $\lambda_1$, respectively. }
    \label{sfig5}
\end{sfigure*}

\begin{sfigure*}
    \centering
    \includegraphics[width= 1.00\textwidth]{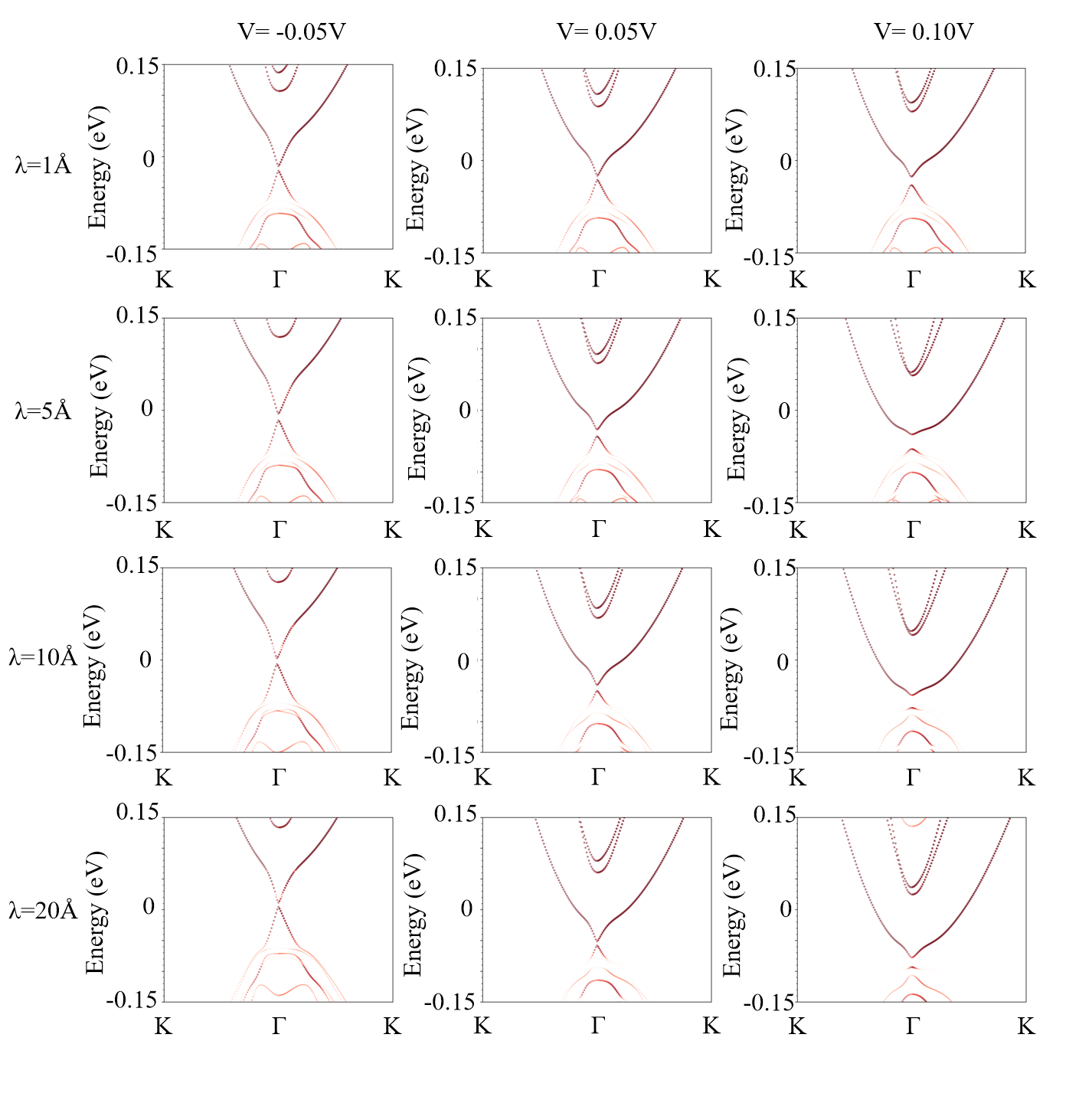}
    \caption{\justifying Similar to SFIG. \ref{sfig5} but for the point-charge-like filed $V_2$.}
    \label{sfig6}
\end{sfigure*}

\begin{sfigure*}
    \centering
    \includegraphics[width= 1.00\textwidth]{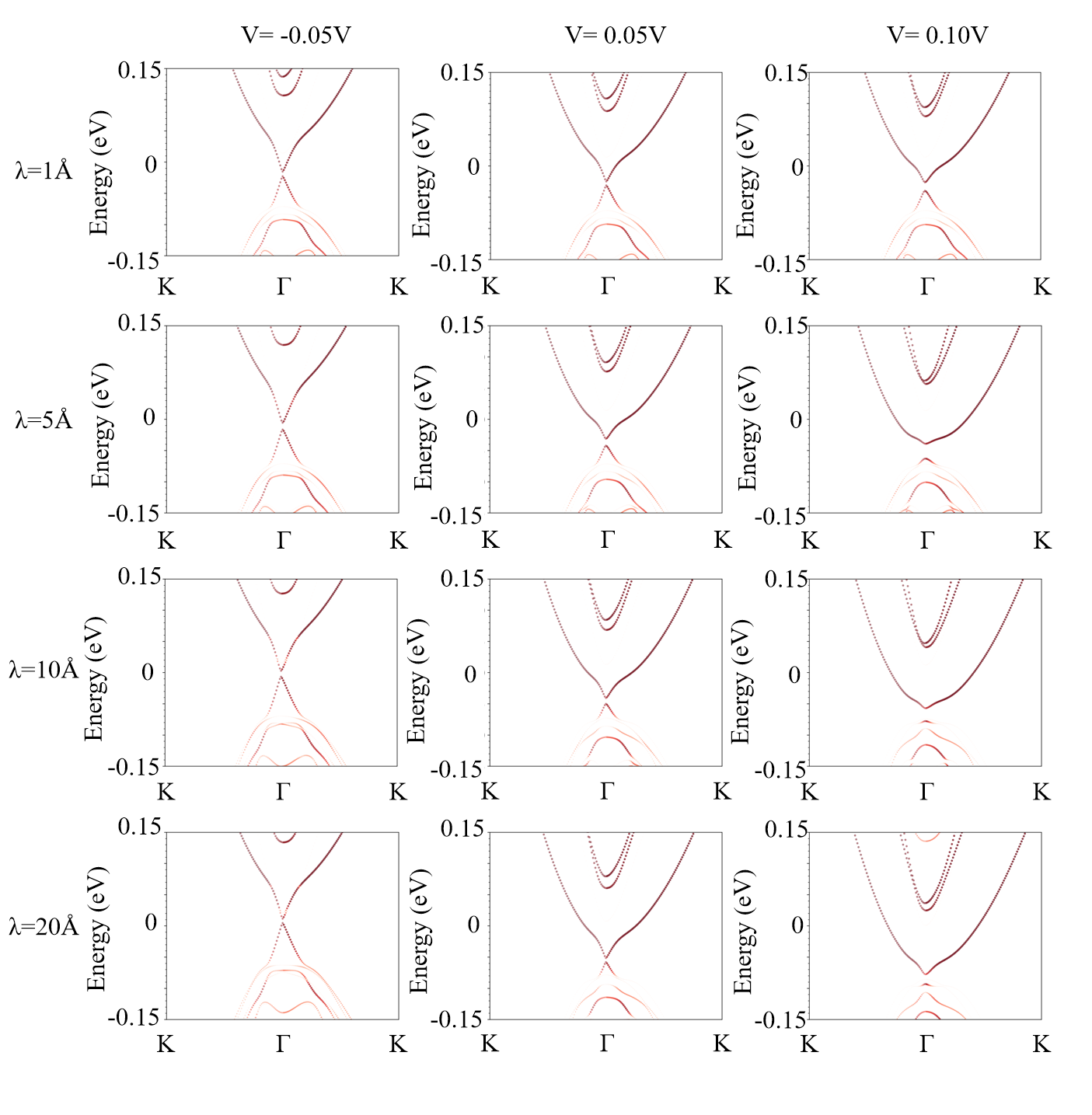}
    \caption{\justifying Similar to SFIG. \ref{sfig5} but for the exponentially decaying filed $V_3$. }
    \label{sfig7}
\end{sfigure*}

\begin{sfigure*}
\centering
\includegraphics[width=1\textwidth]{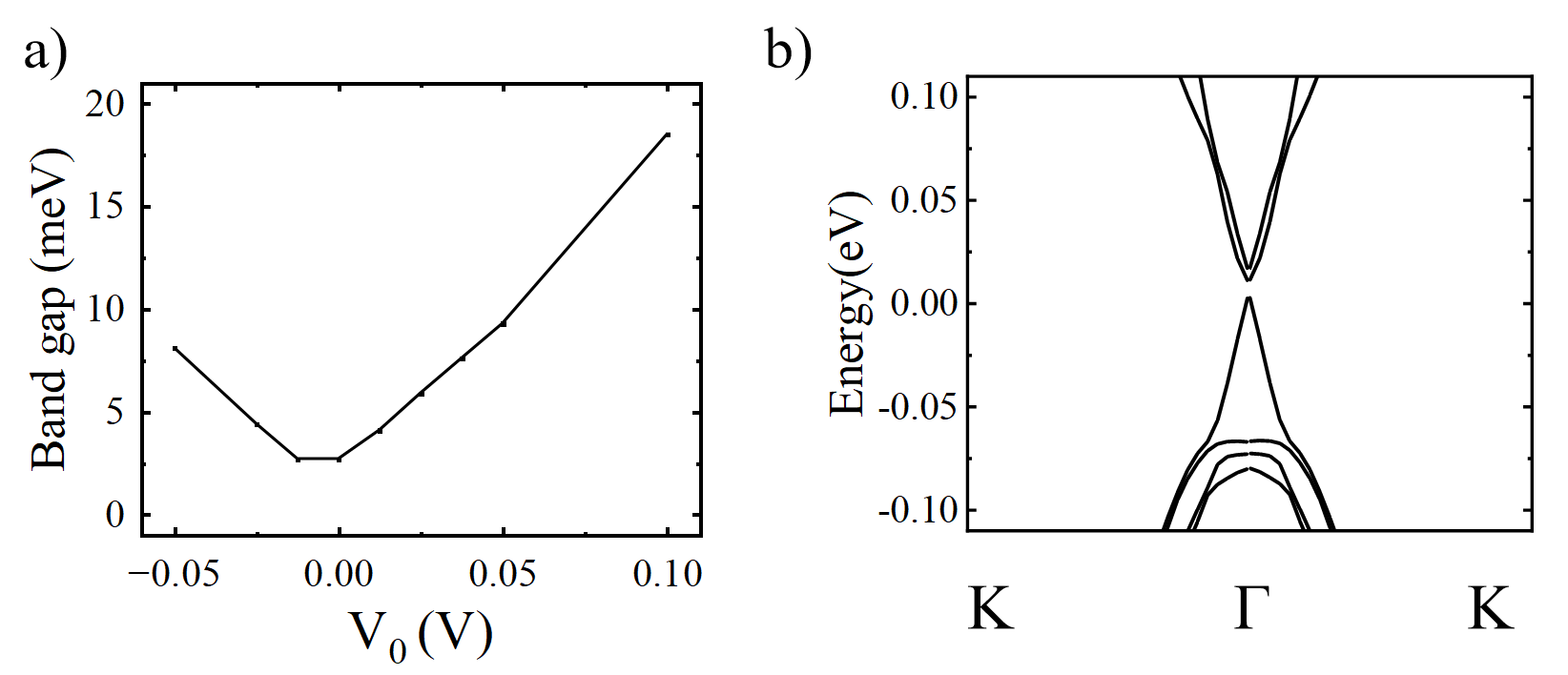}
\caption{\justifying \BLUE{a) Magnetic gap of the 4SL BNN defect structure obtained from DFT calculations. The electric field in the DFT calculations was introduced by applying a homogeneous electric field strength across the slab, which leads to an electrostatic potential difference $V_0$ between the two surfaces. As the thickness of the 4SL slab is approximately equal to $\lambda_1 = 50\,\mathrm{\AA}$, this result is thus directly comparable to the uniform-electric-field ($V_1$) result presented in the main text.
b) Corresponding DFT band structures at a field strength of 0.1 V.
The comparable gap reopening obtained from DFT and tight-binding calculations [Fig. 2a) of the main text] validates the reliability of our model.
}
}
\label{Rfig2}
\end{sfigure*}

\begin{sfigure*}
\centering
\includegraphics[width=1\textwidth]{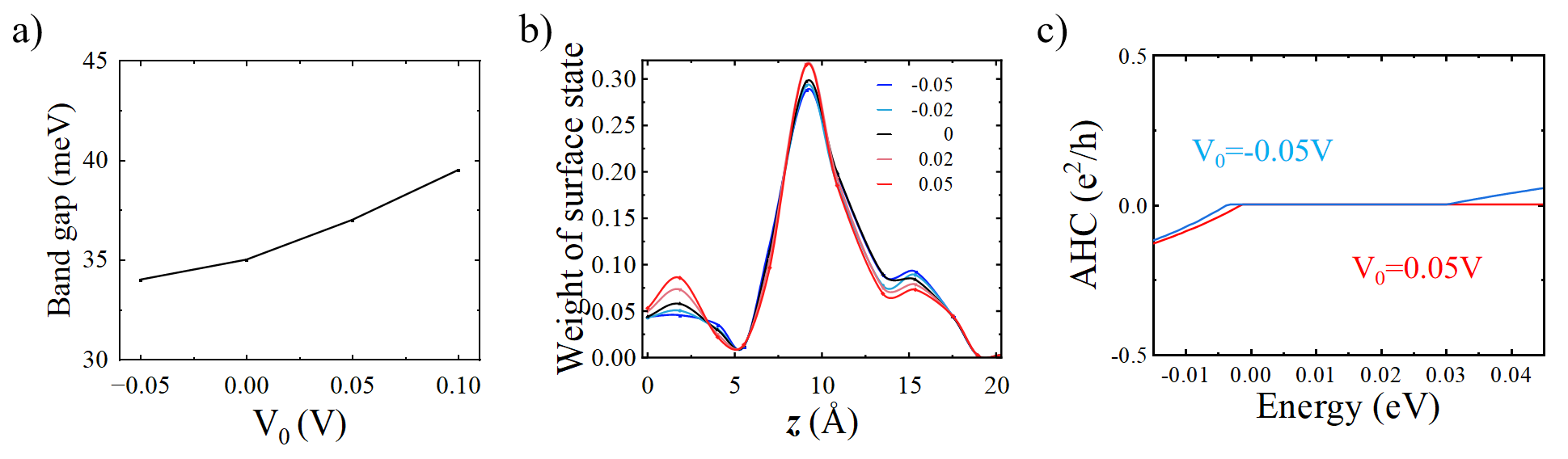}
\caption{ \justifying \BLUE{ (a) Band gap evolution of the 4-SL SNN-defective surface under the exponentially decaying field $V_3$. (b) Distribution of the topological surface states under varying $V_0$. (c) AHC of the 4 SL SNN slab at $V_0 = \pm 0.05$ V.
SNN shows the same qualitative response as BNN in Fig. 2: increasing $V_0$ enhances both the gap and surface-state localization. However, stronger initial localization suppresses gap closing, preventing topology switching within the explored $V_0$ range; a much larger negative $V_0$ would thus be required for the topological transition.
}
}
\label{Rfig4}
\end{sfigure*}

\begin{sfigure*}
\centering
\includegraphics[width=1\textwidth]{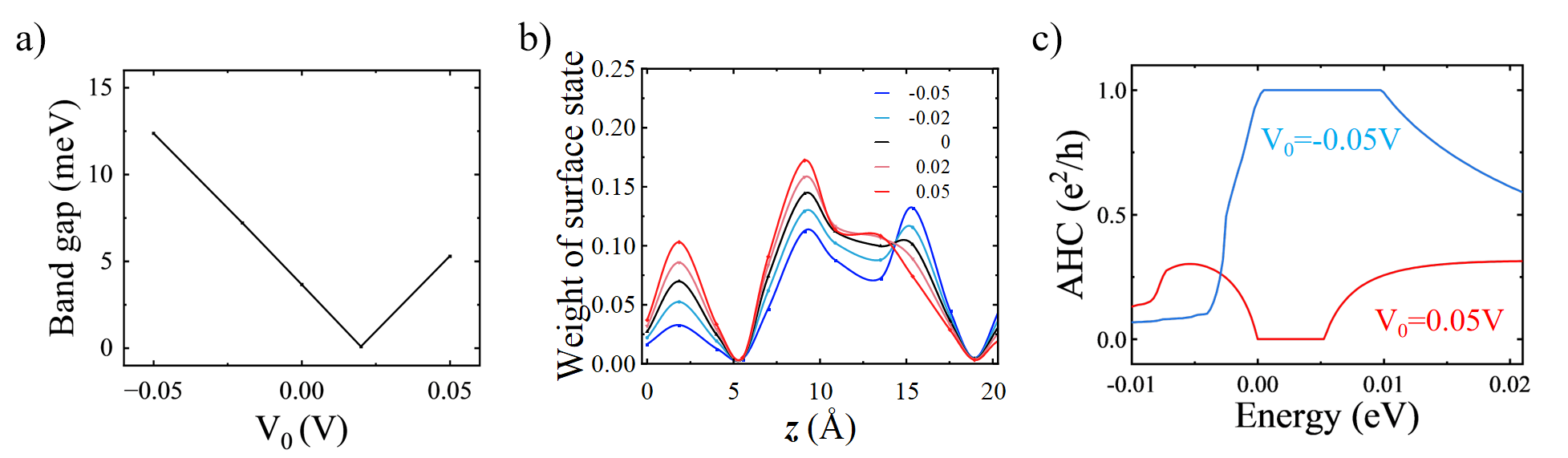}
\caption{ \justifying \BLUE{ a) Magnetic gaps of the V$_{\mathrm{Te^{\prime}}}$ defective surface as a function of $V_0$ under the exponentially decaying field ($V_3$). b) Distribution of the valence band maximum under varied $V_0$. c) AHC of the four-SL slab (bearing V$_{\mathrm{Te^{\prime}}}$ in one surface) under $V_3$ with $V_0 = \pm 0.05$ V. These results are similar to the 4 SL BNN case in Fig. 2 in the main text.}
}
\label{Rfig1}
\end{sfigure*}

\begin{sfigure*}
\centering
\includegraphics[width=0.75\textwidth]{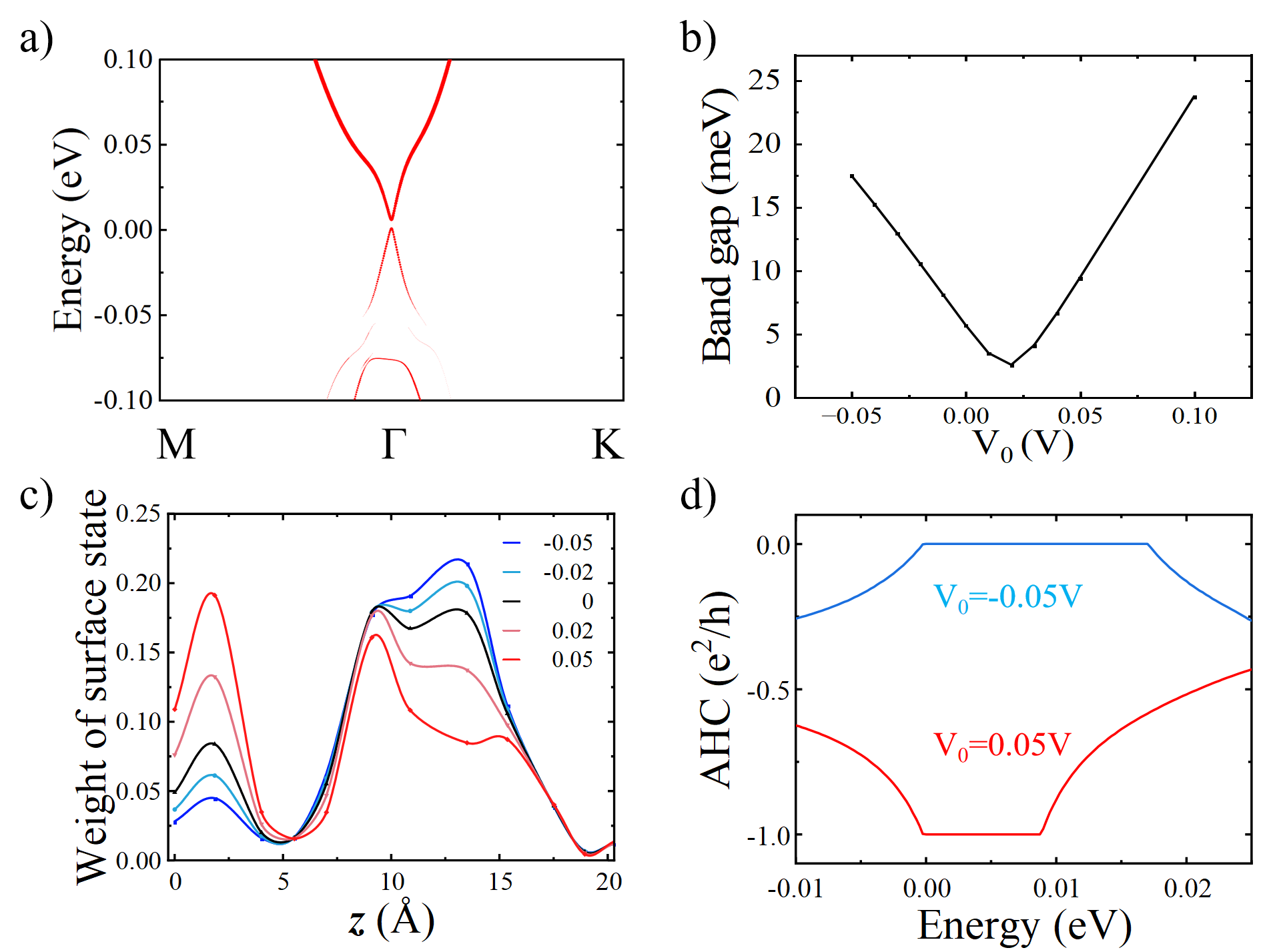}
\caption{ \justifying \BLUE{a) Surface band structure of a 5 SL slab with BNN defects in one surface. 
b) Band gap evolution under the exponentially decaying field $V_3$.
c) Distribution of the topological surface state at different $V_0$.
d) AHC of the 5 SL BNN slab under $V_3$ with $V_0 = \pm 0.05$ V.
These results are consistent with those of the 4 SL slab in Fig. 2, except that the topology of the gap is reversed as expected.
These results demonstrate that the gap size and topology tunability are insensitive to slab thickness.
}}
\label{Rfig5}
\end{sfigure*}

\begin{sfigure*}
\centering
\includegraphics[width= 0.6\textwidth]{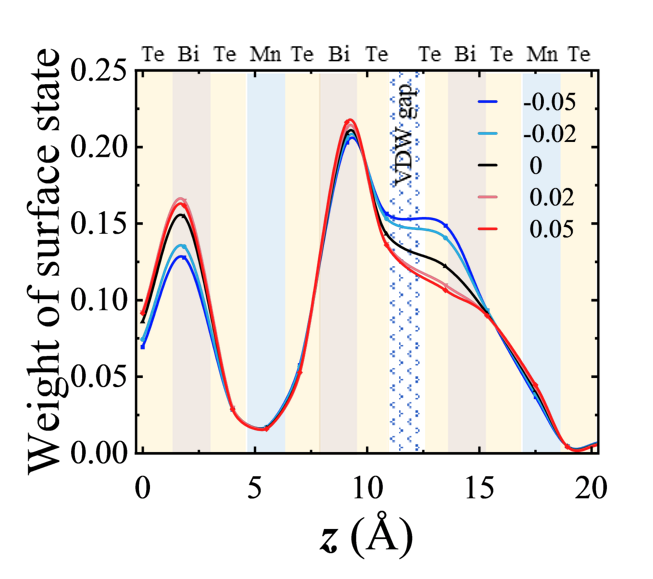}
\caption{\justifying \BLUE{Spatial distribution of the conduction band minimum on the BNN defective surface under the exponentially decaying electric field $V_3$ ($V_0=0.1 $V, $\lambda_3=5$ \AA). These results are complementary to the distribution of the valence band maximum in Fig. 2b) of the main text.}
}
\label{sfig8}
\end{sfigure*}

\begin{sfigure*}
\centering
\includegraphics[width=1\textwidth]{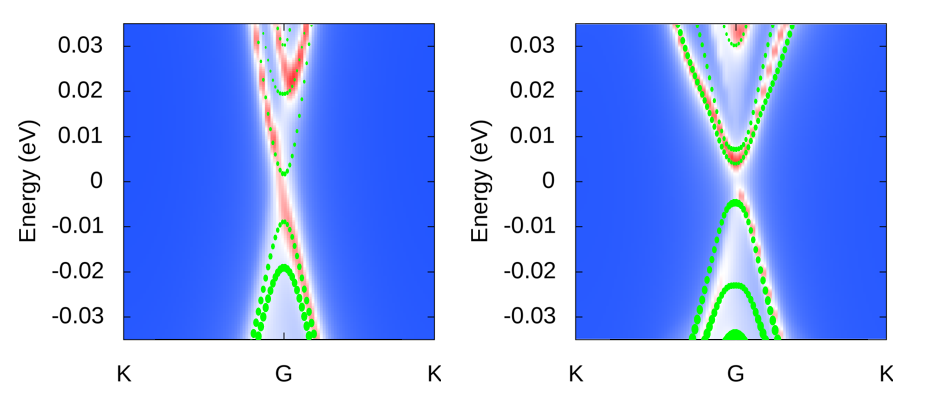}
\caption{ \justifying \BLUE{
Energy dispersions of the [120] edge of the 4-SL BNN-defective slab under an exponentially decaying external potential with (a) negative and (b) positive $V_0$, using $\lambda_3 = 5~\mathrm{\AA}$. Darker red indicates a larger edge-state spectral weight. The green dots represent the slab bands, serving as a guide to the surface gap position.
}}
\label{Rfig8}
\end{sfigure*}

\begin{sfigure*}
\centering
\includegraphics[width= 1.00\textwidth]{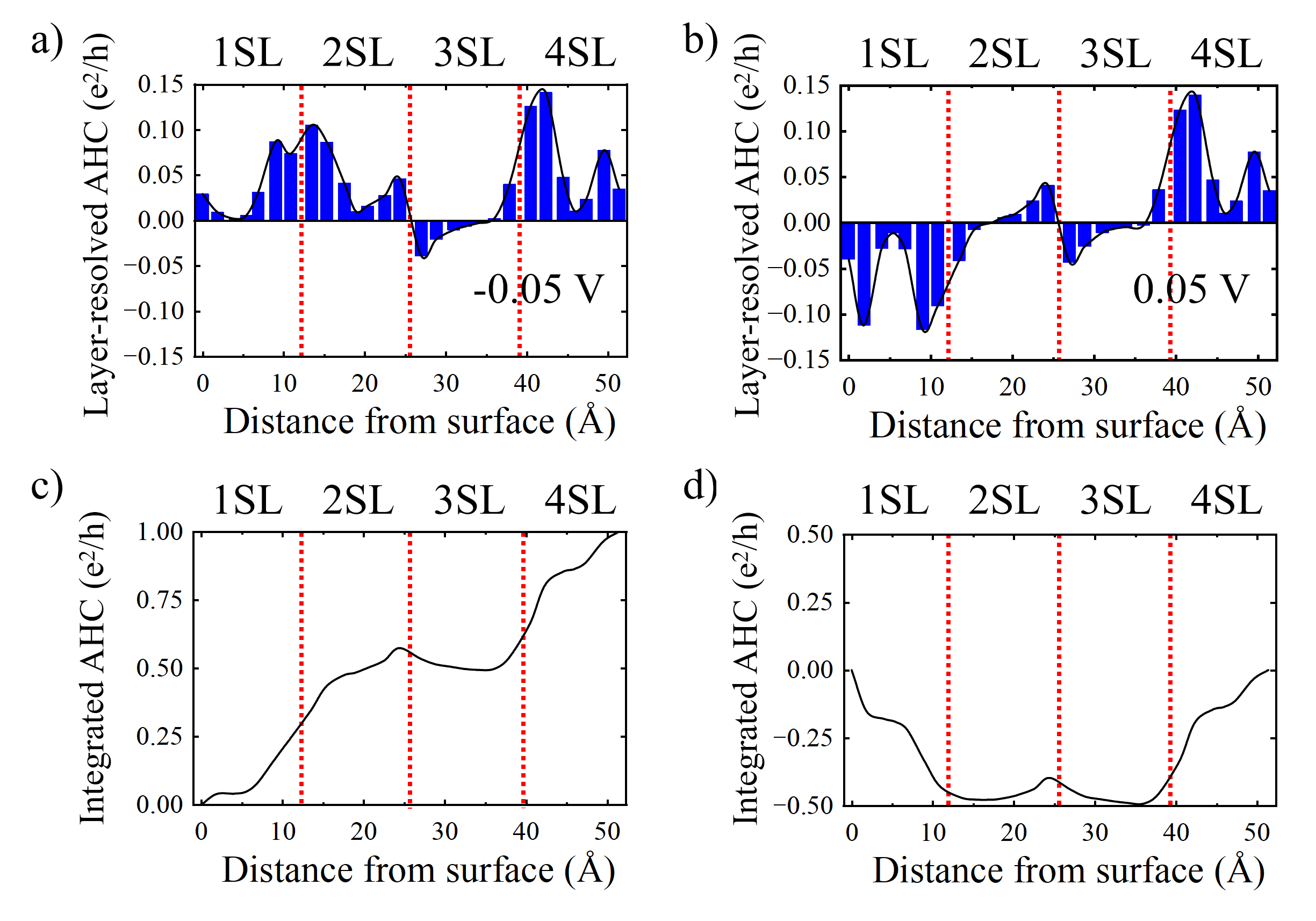}
\caption{\justifying
Layer-resolved anomalous Hall conductivity (AHC) of a four-SL MnBi$_2$Te$_4$ slab with surface BNN co-antisite defects under the exponentially decaying field $V_3$ ($\lambda_3=5$ Å). The distance origin is set at the top (defective) surface, and red dashed lines indicate the vdW gaps. The bottom (pristine) surface is located at about 51 \AA.
Panels (a,b) show the atomic layer-resolved AHC for $V_0=-0.05$ V and $V_0=0.05$ V, respectively; panels (c,d) show the corresponding integrated AHC from the surface. For $V_0<0$ in (a), surface states shift toward the second SL, giving a large positive AHC near the 1SL–2SL vdW gap. For $V_0>0$ in (b), the AHC becomes nearly antisymmetric across the slab with negative values of AHC concentrated in 1SL, yielding zero net AHC for the whole slab in (d).
}
\label{sfig9}
\end{sfigure*}

\begin{sfigure*}
\centering
\includegraphics[width= 1.00\textwidth]{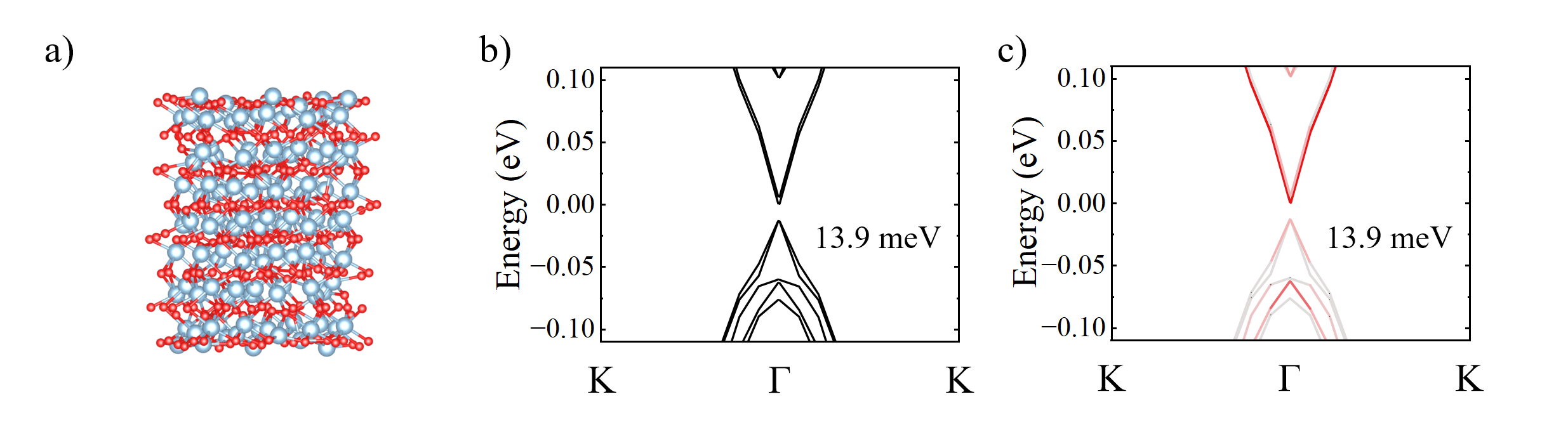}
\caption{\justifying Results of non-stoichiometric AlO$_x$.
a) The relaxed slab structure of the oxygen-rich alumina AlO$_{1.68}$, which is involved in experiments.
b) Raw band structure of the BNN-defective MnBi$_2$Te$_4$-AlO$_x$ heterostructure. 
c) The surface SL projected dispersion in the heterostructure.
}
\label{sfig10}
\end{sfigure*}

\begin{sfigure*}
    \centering
    \includegraphics[width= 1.00\textwidth]{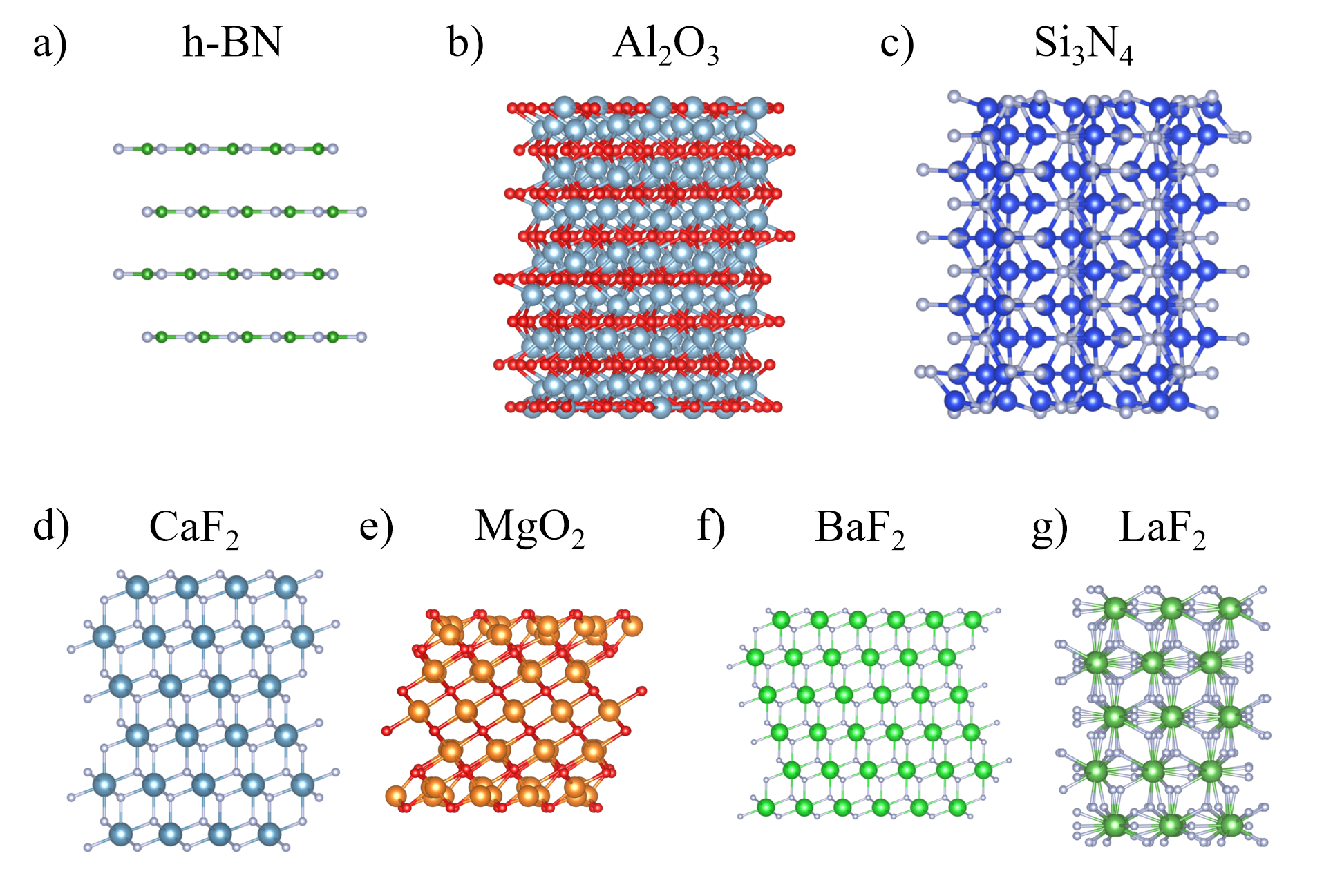}
    \caption{\justifying Slab structures of a) hexagonal BN (h-BN), b) Al$_2$O$_3$, c) Si$_3$N$_4$, d) CaF$_2$, e) MgO$_2$, f) BaF$_2$, and g) LaF$_3$ used in the calculations. All surfaces are fully relaxed, where the MgO surface shows reconstruction. All configurations are insulating.}
    \label{sfig11}
\end{sfigure*}

\begin{sfigure*}
\centering
\includegraphics[width=1\textwidth]{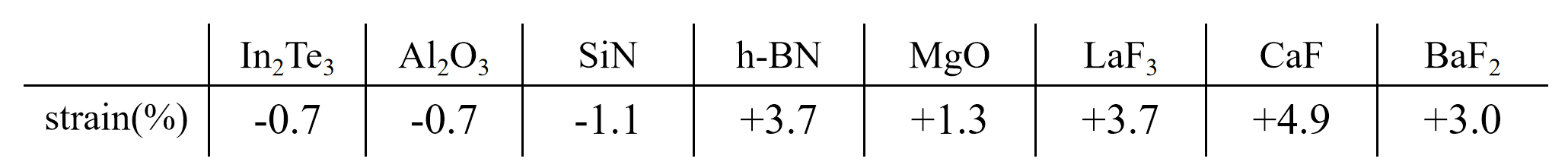}
\caption{\justifying \BLUE{A summary of strain applied to capping layers in heterostructure calculations (in unit of percent \%). Positive values mean tensile strain, while negative values mean compressive strain.}
}
\label{Rtable3}
\end{sfigure*}

\begin{sfigure*}
    \centering
    \includegraphics[width= 1.00\textwidth]{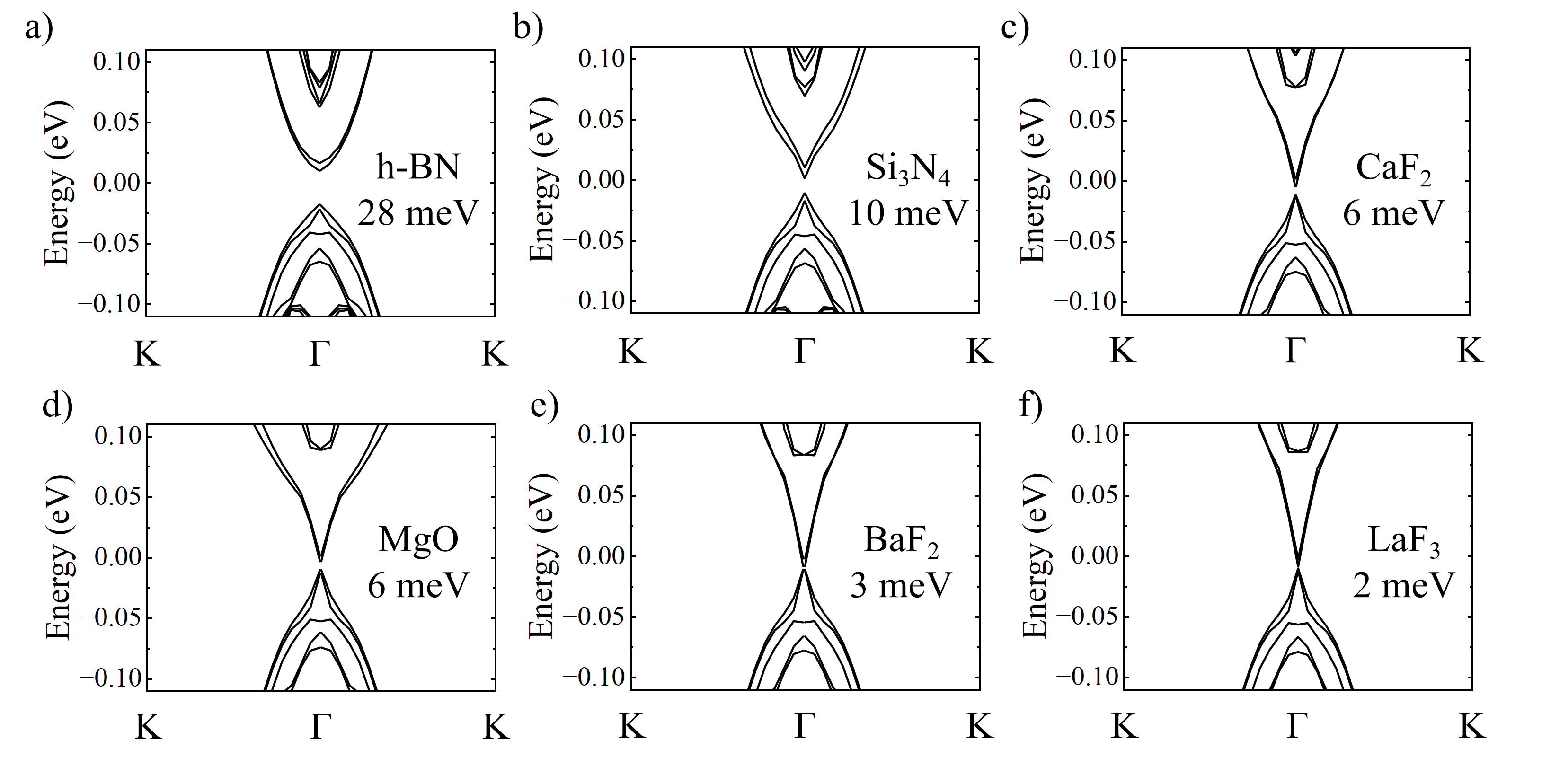}
    \caption{\justifying Band structures of heterostructures formed by the BNN-defective MnBi$_2$Te$_4$ and the insulating layers in SFIG. \ref{sfig11} at the vdW spacing of $2.8\,$\AA. Plots are ordered by descending gap value: a) h-BN, b) Si$_3$N$_4$, c) CaF$_2$, d) MgO$_2$, e) BaF$_2$, and f) LaF$_3$. Results for the MnBi$_2$Te$_4$$-$Al$_2$O$_3$ are shown in Fig. 4b) in the main text.}
    \label{sfig12}
\end{sfigure*}

\begin{sfigure*}
\centering
\includegraphics[width= 1.00\textwidth]{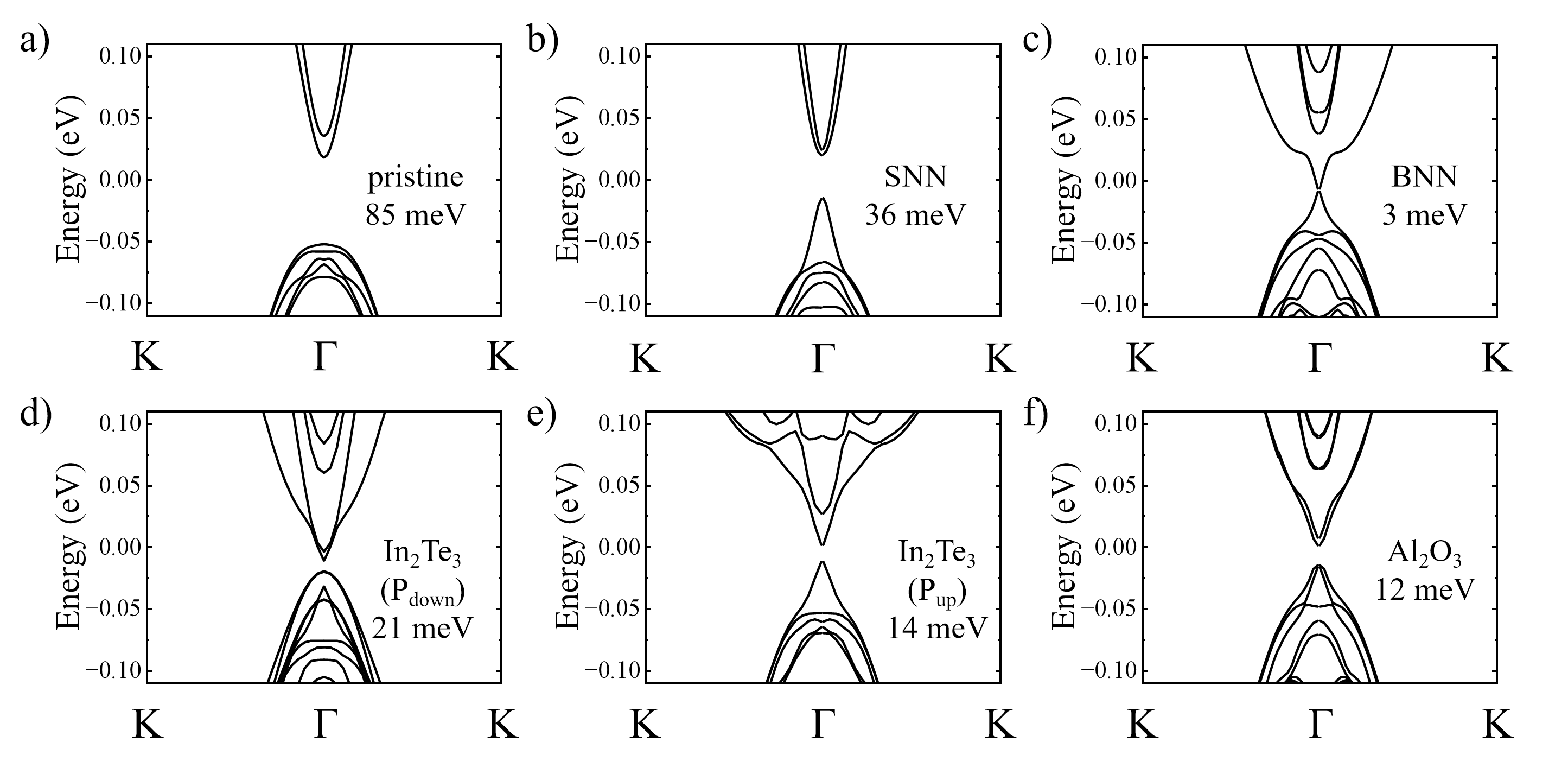}
\caption{\justifying Raw band structures of a) pristine surface, b) SNN defective surface, c) BNN defective surface, d) BNN-defective MnBi$_2$Te$_4$-In$_2$Te$_3$ heterostructure with P$_{down}$ In$_2$Te$_3$, e) BNN-defective MnBi$_2$Te$_4$-In$_2$Te$_3$ heterostructure with P$_{up}$ In$_2$Te$_3$, and f) BNN-defective MnBi$_2$Te$_4$-Al$_2$O$_3$ heterostructure. The corresponding surface-projected band structures are shown in the main text. The value of the surface Dirac point gap is also shown in each panel.}
\label{sfig13}
\end{sfigure*}

\clearpage

%